\documentclass[11pt]{article}
\newtheorem{lemma}{Lemma}[section]

\newtheorem{theorem}{Theorem}[section]

\usepackage[font=small,labelfont=bf]{caption}
\usepackage{amssymb}
\usepackage{amsmath}
\usepackage{epsfig}
\usepackage{graphics}
\date{}
\title{A new estimator for Weibull distribution parameters: Comprehensive comparative study for Weibull Distribution\\
\bigskip
\small{Sahar Sadani}\\
Email: s.sadani@gu.ac.ir\\
Department of Statistics, Faculty of Science, Golestan University, Gorgan, Iran.\\
\bigskip
\small{Kamel Abdollahnezhad}\\
Email: k.abdollahnezhad@gu.ac.ir\\
Department of Statistics, Faculty of Science, Golestan University, Gorgan, Iran.\\
\bigskip
\small{Mahdi Teimouri}\\
Email: teimouri@aut.ac.ir\\
Department of Mathematics and Statistics, Faculty of Science and Engineering, Gonbad Kavous University, Gonbad Kavous, Iran.\\
\bigskip
\small{Vahid Ranjbar}\\
Email: v.ranjbar@gu.ac.ir\\
Department of Statistics, Faculty of Science, Golestan University, Gorgan, Iran.}
\begin{document}
\maketitle{}
\noindent{}
%\newpage{}
%{\center{\LARGE{A new estimator for Weibull distribution parameters: \\Comprehensive comparative study for Weibull Distribution}}}
\bigskip\\
{\Large{\bf{Abstract}}}:\\
~~Weibull distribution has received a wide range of applications in engineering and science. The utility and usefulness of an estimator is highly subject to the field of practitioner's study. In practice users looking for their desired estimator under different setting of parameters and sample sizes. In this paper we focus on two topics. Firstly, we propose $U$-statistics for the Weibull distribution parameters. The consistency and asymptotically normality of the introduced $U$-statistics are proved theoretically and by simulations. Several of methods have been proposed for estimating the parameters of Weibull distribution in the literature. These methods include: the generalized least square type 1, the generalized least square type 2, the $L$-moments, the Logarithmic moments, the maximum likelihood estimation, the method of moments, the percentile method, the weighted least square, and weighted maximum likelihood estimation. Secondary, due to lack of a comprehensive comparison between the Weibull distribution parameters estimators, a comprehensive comparison study is made between our proposed $U$-statistics and above nine estimators. Based on simulations, it turns out that the our proposed $U$-statistics show the best performance in terms of bias for estimating the shape and scale parameters when the sample size is large.
\section{Introduction}
\setcounter{equation}{0}
The Weibull distribution is one of the most commonly used distributions with a wide range of applications in some study fields such as:
chemical engineering (\cite{chiang2004examination}, \cite{kuo2009new}, and \cite{wood2005temporal}),
ecology \cite{stankova2010modeling},
electrical engineering (\cite{genc2005estimation} and \cite{pascual2006morchella}),
food industry \cite{corzo2008weibull},
mechanical engineering (\cite{raghunathan2002studies} and \cite{lavanya2016fast}),
telecommunications (\cite{surendran2014parameter} and \cite{buller2013statistical}),
wireless communications \cite{noga2010overview},
economic (\cite{nadarajah2006modified} and \cite{diaconu2009weibull}),
civil engineering (\cite{muraleedharan2007modified} and \cite{arenas2010optimum}), and
seismology \cite{hasumi2009weibull}. For further details on applications of the Weibull distribution, we refer the readers to
Meeker and Escobar (1998), Murthy {\it et al.} (2004), and Dodson (2006). However, a comprehensive study has not been performed to compare the estimators.
All comparative studies, to the best of our knowledge, have been devoted to compare the performance of the MLE with the estimators of another class. For example, Kanter (2015) made a comparison between least square estimators and the MLEs. The bias of the MLE for the Weibull distribution has been studied by Ross (1996), Watkins (1996) and Montanari {\it et al.} (1997). Seki and Yokoyama (1996) made a comparison between the MLE and a bootstrap estimator. Zhang {\it et al.} (2007) compared the estimation methods based on the Weibull probability plot. We also refer the readers to \cite{gebizlioglu2011comparison}, \cite{genschel2010comparison}, \cite{hossain2003comparison}, \cite{mohan2013comparison}, \cite{teimouri2013comparison}, and references therein. This is while estimators may have different appeals to different users. For example, the maximum likelihood estimator (MLE) that has attractive properties is biased and has not closed-form expression. This is while the practitioners from some fields may looking for an estimator that is unbiased or has closed-form expression.
Also, user may prefer to use an estimator which works satisfactorily with sample of small size.
Hence, a comparative study is needed to compare the performance of the known estimators under different situations. In this paper, we perform a comprehensive comparison study between ten class of estimators including: the generalized least square type 1 (GLS1), the generalized least square type 2 (GLS2), the $L$-moments (LM), the Logarithmic moments (MLM), the maximum likelihood estimation (MLE), the method of moments (MM), the percentile method (PM), the $U$-statistic, the weighted least square (WLS), and
weighted maximum likelihood estimation (WMLE).
\par The structure of the paper is as follows. In Section 2, we derive $U$-statistic for the shape and scale parameters of the Weibull distribution.
The known estimation methods are reviewed briefly in Section 3.
A comparison between proposed estimator and the known ones as well as a real data illustration are given in Section 4.
\section{$U$-statistics for the Weibull distribution parameters}
\setcounter{equation}{0}
The probability density function (pdf) and cumulative distribution function (cdf) of two-parameter
Weibull distribution are, respectively, given by (Nelson, 1982; Johnson {\it et al.}, 1994; Dodson, 2006):
\begin{align}
f_{X}(x)&=\frac{\alpha }{\beta}
\left(\frac{x}{\beta } \right)^{\alpha -1} \exp
\biggl\{-\left(\frac{x}{\beta } \right)^{\alpha } \biggr\},\label{pdf}\\
F_{X}(x)&=1- \exp
\biggl\{-\left(\frac{x}{\beta } \right)^{\alpha } \biggr\},\label{cdf}
\end{align}
for $x>0$, $\alpha > 0$ and $\beta >0$. Here, $\alpha$ and $\beta$ are known as the shape and scale parameters.
In the following we give $U$-statistics for the shape and scale parameters of the Weibull distribution. For this, a lemma given by the following is necessary. Hereafter, we write ${\cal{W}}(\alpha,\beta)$ to denote a Weibull distribution with pdf given in (\ref{pdf}).
\begin{lemma} Suppose $X_1,X_2 \mathop \sim \limits^{iid} {\cal{W}}(\alpha,\beta)$. Then
\begin{align}\label{rep}
\min\{X_1,X_2\}\mathop= \limits^{d} 2^{-\frac{1}{\alpha}}X_1.
\end{align}
\end{lemma}
{\bf{Proof:}} Define $Y=\min\{X_1,X_2\}$.
%It is known that
%\begin{align*}
%F_Y(y)=P(Y\leq y)=1-\left[1-F_X(y)\right]^2,
%\end{align*}
%where $F_X(.)$ is the cdf of Weibull distribution given in (\ref{cdf}).
It follows that
\begin{align}\label{Fy}
F_Y(y)=1-\exp \biggl\{-2\left(\frac{y}{\beta } \right)^{\alpha } \biggr\},~~~y>0.
\end{align}
On the other hand, define $Z=2^{\frac{1}{\alpha}}X_1$. We have,
\begin{align}\label{Fz}
F_Z(y)=P(Z\leq y)=1-\exp \left\{-\left(\frac{y}{2^{-\frac{1}{\alpha}}\beta } \right)^{\alpha}\right\}.
\end{align}
Comparing the right-hand sides of (\ref{Fy}) and (\ref{Fz}), it turns out that $F_Y(y)=F_Z(y)$; for $y>0$, and so the result follows.
\begin{theorem}
Suppose $x_1,x_2,\dots,x_n$ are $n$ independent realizations from Weibull distribution with pdf given in (\ref{pdf}). Then,
\begin {enumerate}
\item \begin{equation*}\label{ustat}
U_{\alpha}={\dbinom {n}{2}}^{-1}\sum_{1\leq i<j \leq n} H_{1}(x_{i},x_{j}),
\end{equation*}
in which
\begin{equation*}
H_{1}(x_{i},x_{j})=\frac{\log \min\{x_{i},x_{j}\}}{\log 2}-\frac{\log x_{i}+\log x_{j}}{2\log 2},
\end{equation*}
is $U$-statistic for $1/\alpha$.
\item \begin{equation*}\label{ustat}
U_{\sigma}={\dbinom {n}{2}}^{-1}\sum_{1\leq i<j \leq n} H_{2}(x_{i},x_{j}),
\end{equation*}
in which
\begin{equation*}
H_{2}(x_{i},x_{j})=\biggl(1+\frac{\psi(1)}{\log 2}\biggl)\frac{\log x_{i}+\log x_{j}}{ 2}-\frac{\psi(1)}{\log 2}\log \min\{x_{i},x_{j}\},
\end{equation*}
where $\psi(1)=-0.57721566$ is $U$-statistic for $\log \sigma$.
\end{enumerate}
\end{theorem}

{\bf{Proof:}}
\begin{enumerate}
\item
By applying log-transformation to the both sides of (\ref{rep}), we have
\begin{eqnarray}\label{identity1}
\frac{1}{\alpha}=\frac{\log \min\{X_{1},X_{2}\}-\log {{X}}_{1}}{\log 2}.
\end{eqnarray}
The right-hand side of (\ref{identity1}) can be used to construct a symmetric kernel as
\begin{eqnarray}\label{kernel1}
H_1(x_{1}, x_{2})=\frac{\log x_{1}+\log x_{2}}{2\log 2}-\frac{\log \min\{x_{1}, x_{2}\}}{\log 2}.
\end{eqnarray}
It is easy to see that $ \text{E}\bigl(H_1({X}_{1},{X}_{2})\bigr)=1/\alpha$. To guarantee the asymptotic normality of the introduced $U$-statistics for $1/\alpha$ with kernel (\ref{kernel1}), it is necessary to show that $
\text{Var}\bigl(E\left(H_1({X}_{1},{X}_{2})\big|{X}_{1}\right)\bigr)$ is finite. For this, it suffices to show that
$\text{Var}\bigl(H_1({{X}}_{1},{{X}}_{2})\bigr)$ is finite. To begin, we note that $X, X_1, X_2 \mathop \sim \limits^{iid} {\cal{W}}(\alpha,\beta)$. Since $\min\{X_1,X_2\}\mathop= \limits^{d} 2^{-\frac{1}{\alpha}}X_1$, it follows that
\begin{align}
\text{Var} (\log X)&=\text{Var} \left(\log \min\{X_{1},X_{2}\}\right)\label{cov1}.
\end{align}
Also, suppose $X$, $Y$, and $Z$ are given arbitrary random variables. Generally, we cannot conclude that if $X\mathop= \limits^{d}Y$, then $\text{Cov}\bigl(X,Z\bigr)=\text{Cov}\bigl(Y,Z\bigr)$. But, here, elementary statistical manipulations reveal that if we define $X=\log \min\{X_{1},X_{2}\}$, $Y=\log X_1$, and $Z=\log X_1+\log X_2$, then
\begin{align}
\text{Cov}\left(\log \min\{X_{1},X_{2}\}, \log X_1+ \log X_2\right)&=\text{Cov}\left(\log X_1, \log X_1+\log X_2\right).\label{cov2}
\end{align}
Now, using (\ref{cov1}), we can write
\begin{align} \label{varh}
\text{Var}\bigl(H_1({{X}}_{1},{{X}}_{2})\bigr)=\frac{\text{Var} \log X}{\log^{2} 2}+\frac{\text{Var} \log X}{2\log^{2} 2}-\frac{\text{Cov}\bigl(\log \min\{X_{1},X_{2}\}, \log X_1+ \log X_2\bigr)}{\log^{2} 2}.
\end{align}
Applying property (\ref{cov2}) to the right-hand side of (\ref{varh}), we obtain
\begin{eqnarray*}
\text{Var}\bigl(H_1({{X}}_{1},{{X}}_{2})\bigr)\leq\frac{\text{Var}(\log X)}{2\log^{2} 2}.
\end{eqnarray*}
It is easy to check that
$\text{Var} (\log X)= \psi(1,1)/\alpha^2$
where $\psi(n,x)=\partial^n \psi(x)/\partial x^n$ and $\psi(x)=\partial \log \Gamma(x)/\partial x$. The asymptotic normality of $U_{\alpha}$ follows since
\begin{eqnarray*}
\text{Var}\bigl(H_1({{X}}_{1},{{X}}_{2})\bigr)\leq\frac{\psi(1,1)}{2\alpha^2\log^{2} 2}<\infty.
\end{eqnarray*}
\item It is not hard to check that $\text{E} (\log X)=\log \beta+\frac{\psi(1)}{\alpha}$ where $\psi(1)=-0.5772157$.
Define $ H_2(x_{1},x_{2}) $ as
\begin{align}\label{kernel2}
H_2(x_{1},x_{2}) &= \frac {\log x_1+ \log x_2}{2}-\psi(1) H_1(x_{1},x_{2})\nonumber\\
&= \frac {\log x_1+ \log x_2}{2}\biggl(1-\frac{\psi(1)}{\log 2}\biggr)+\frac{\psi(1)}{\log 2}\log \min\{x_{1},x_{2}\}.
\end{align}
It is easy to see that $ \text{E}\bigl(H_1({{X}}_{1},{{X}}_{2})\bigr)=\log \beta$. Asymptotic normality of the introduced $U$-statistics for $\log \beta$ with kernel (\ref{kernel2}) holds if we prove $\text{Var}\bigl(E\left(H_2({X}_{1},{X}_{2})\big|{X}_{1}\right)\bigr)<\infty$ or equivalently $\text{Var}\bigl(H_1({{X}}_{1},{{X}}_{2})\bigr)<\infty$. We eliminate the proof since kernels $H_1(x_{1}, x_{2})$ and $H_2(x_{1}, x_{2})$ have similar structure.
\end{enumerate}
\section{Known estimators for the Weibull distribution}
\setcounter{equation}{0}
Here, we review briefly almost all of known estimation methods for the Weibull distribution.
\subsection{Maximum likelihood estimation (MLE)}
There is no closed-form expression for MLEs of the Weibull distribution parameters. It is asymptotically normal and efficient for large sample sizes. Many attempts have been made to compute or modify the MLEs of the Weibull distribution parameters. Cohen and Whitten (1982) considered a modified MLE involving complicated numerical computations. Dodson (2006) derived the MLE for the shape parameter graphically. The MLE of the shape parameter is computed as the root of the equation, see \cite{norman1994continuous}
\begin{eqnarray*}
\displaystyle
\frac{n}{\alpha}-\sum_{i=1}^{n}\log x_{i}-n \log \beta+ \sum_{i=1}^n \biggl(\frac{x_i}{\beta}\biggr)^{\alpha} \log \biggl(\frac{x_i}{\beta}\biggr),
\end{eqnarray*}
and the MLE of the scale parameter is given by
\begin{eqnarray*}
\widehat {\beta}_{MLE}=\left(\frac{{\sum_{i=1}^n {x_{i}^{\alpha}}}}{n} \right)^{\frac{1}{\alpha}}.
\end{eqnarray*}
It can be seen that $\widehat {\beta}_{MLE}$ depends on $\alpha$ and also that $\widehat{\alpha }_{MLE}$ must be computed numerically.
\subsection{Weighted Maximum likelihood (WMLE)}
It is known that MLEs are generally biased. To reduce the bias rate in the case of the Weibull distribution, the weighted maximum likelihood estimators (WMLE) have been proposed in \cite{jacquelin1993generalization}. Suppose $x_1, x_2, \ldots, x_n $ is a random sample from cdf given in (\ref{cdf}), then the WMLEs of the shape and scale parameters are given by
\begin{align*}
{\hat{\alpha}}_{WMLE}&=\arg\min\limits_{\alpha} ~\biggl(\frac{W_2}{\alpha}+\frac{1}{n}\log x_i-\frac{\sum_{i=1}^n x_{i}^{\alpha} \log x_i}{\sum_{i=1}^n x_{i}^{\alpha}}\biggr)^2,\nonumber\\
{\hat{\beta}}_{WMLE}&=\biggl(\frac{1}{nW_1}\sum_{i=1}^n x_{i}^{\alpha}\biggr)^{\frac{1}{\alpha}},\nonumber
\end{align*}
where the weights $W_1$ and $W_2$ are given by
\begin{align*}
W_1&=-\frac{1}{n}\sum_{i=1}^{n}\log (1-F(X_i)),\nonumber\\
W_2&=\frac{\sum_{i=1}^{n}\log \bigl(1-F(X_i)\bigr) \log\bigl(-\log (1-F(X_i))\bigr)}{\sum_{i=1}^{n}\log (1-F(X_i))}-\frac{1}{n}\sum_{i=1}^{n}\log\bigl(-\log (1-F(X_i))\bigr).\nonumber
\end{align*}
Although the sampling distribution of the $W_1$ is gamma with shape parameter $n$ and scale parameter $1/n$, but the sampling distribution of the $W_2$ is not known. In practice, both of random variables $W_1$ and $W_2$ are replaced by their central quantities such as mean, median, or geometric mean. Here, we use the median of $W_1$ and $W_2$ since they yield the best performance, see \cite{cousineau2009nearly}. For this, in a comprehensive Monte Carlo simulation, we derive the median of $W_1$ and $W_2$ for different levels of $\alpha$ (from 0.5 to 5 by 0.2) and small sample size $n$ (including 5, 10, 15, 30, 50, 100, \dots, 200). We note that as $n$ tends to $\infty$, both WMLE and MLE approaches give the same results.
\subsection{Generalized and weighted least square (GLS and WLS)}
The parameter estimation using least square approach is common in the statistical literature. For Pareto, log-logistic and Weibull distributions we refer the readers to \cite{kantar2015generalized}, \cite{hung2001weighted}, \cite{lu2004note}, \cite{zhang2008weighted}, \cite{van2012parameter}, and \cite{zhang2007study}. Suppose $x_{(1)} \leq x_{(2)}\leq \dots \leq x_{(n)}$ are the ordered realizations from Weibull distribution with pdf given in (\ref{cdf}). We can see that the following regression model holds.
\begin{align}\label{reg1}
y_{(i)}=\log \beta+\frac{1}{\alpha}\log\bigl(-\log (1-F(x_{(i)}))\bigr),
\end{align}
for $i=1,\dots,n$ where $y_{(i)}=\log x_{(i)}$. The quantity $F(x_{(i)})$, in the right-hand side of regression model (\ref{reg1}), is replaced by $\frac{i}{n+1}$ or $\frac{i-0.3}{n+0.4}$, see \cite{ tiryakiouglu2007estimating}, \cite{van2012parameter}, and \cite{kantar2015robust}. Since the sample $x_{(i)}$ is ordered, the dependent variable $y_{(i)}$ is also ordered. Therefore the variance of dependent variable is not of the form $\sigma^2$I, see \cite{kantar2015generalized}. To tackle this issue the generalized least square (GLS) technique is proposed, see \cite{engeman1982generalized}. The GLS estimate, i.e., $\hat{\boldsymbol{\beta}}_{GLS1}=(\log \hat{\beta} ,1/\hat{\alpha})^T$ is given by
\begin{align}\label{gls1}
\hat{\boldsymbol{\beta}}_{GLS1}=\bigl(X^TV^{-1}X\bigr)X^TV^{-1}Y,
\end{align}
where $Y=(\log x_{(1)}, \log x_{(2)},\dots, \log x_{(n)})^T$,
\begin{align*}
X=\begin{pmatrix}
1&\log\bigl(-\log (1-\hat{F}(x_{(1)}))\bigr)\\
\vdots&\vdots\\
1&\log\bigl(-\log (1-\hat{F}(x_{(n)}))\bigr)
\end{pmatrix},
\end{align*}
and
\begin{align*}
V=\begin{pmatrix}
v_{11}&\dots&v_{1n}\\
\vdots&\vdots&\vdots\\
v_{n1}&\dots&v_{nn}
\end{pmatrix},
\end{align*}
for
\begin{align*}
v_{ij}=\frac{i}{(n+1-i)}\frac{1}{\log (n+1-i)-\log (n+1)}\frac{1}{\log (n+1-j)-\log (n+1)}; ~~~~~~ i\leq j.
\end{align*}
The second type of GLS estimate, i.e.,
\begin{align}\label{gls2}
\hat{\boldsymbol{\beta}}_{GLS2}=\bigl(Z^TV^{-1}X\bigr)Z^TV^{-1}Y,
\end{align}
can be constructed if we replace $X$ with $Z$ as
\begin{align*}
Z=\begin{pmatrix}
1&\log\bigl(-\log (1-\hat{F}(x_{(1)}))\bigr)-0.5-\frac{\log (1-\hat{F}(x_{(1)}))}{\bigl((1-\hat{F}(x_{(1)}))\log (1-\hat{F}(x_{(1)}))\bigr)^2}\\
\vdots&\vdots\\
1&\log\bigl(-\log (1-\hat{F}(x_{(n)}))\bigr)-0.5-\frac{\log (1-\hat{F}(x_{(n)}))}{\bigl((1-\hat{F}(x_{(n)}))\log (1-\hat{F}(x_{(n)}))\bigr)^2}
\end{pmatrix}.
\end{align*}
We note that $\hat{\boldsymbol{\beta}}_{GLS2}=(\log \hat{\beta} ,1/\hat{\alpha})^T$ and $\hat{F}(x_{(i)})=\frac{i}{n+1}$. The weighted least square (WLS) estimate are also given by
\begin{align}\label{wls}
\hat{\boldsymbol{\beta}}_{WLS}=\bigl(X^TW^{-1}X\bigr)X^TW^{-1}Y,
\end{align}
where $\hat{\boldsymbol{\beta}}_{WLS}=(\log \hat{\beta} ,1/\hat{\alpha})^T$ and $W$ is a diagonal matrix whose entries are $v_{11},\dots,v_{nn}$, see \cite{kantar2015generalized}.
\subsection{$L$-moment (LM)}
The $L$-moments have their origin in works by Hosking (1990) and Elamir and Seheult (2003).
% This approach has been applied for different aims including:
%statistical inference for generalized Rayleigh and the exponential distributions, see \cite{kundu2005generalized} and \cite{abdul2007moments},
%characterization of distributions (hosking2006characterization; \cite{karvanen2008characterizing},
%dynamic quantile models \cite{gourieroux2008dynamic},
%inference of the quantile mixtures \cite{karvanen2006estimation},
%hedge funds modeling \cite{darolles2009performance}, and
%assessing symmetry \cite{thomas2009assessing}.
By equating the sample $L$-moment to the population counterpart gives the $L$-moment estimate.
The $r$-{th} $L$-moment of Weibull distribution with pdf (\ref{pdf}) is given by:
\begin{eqnarray*}
\mu_{r}^{L}= \frac{\beta}{r} \Gamma \left( \frac{1}{\alpha }+1 \right)
\sum\limits_{k=0}^{r-1} {(-1)^k} C_k^{r-1}
(r-k)C_{r-k}^r \sum\limits_{j=0}^{r-k-1}
C_j^{r-k-1} \frac{\displaystyle (-1)^j}{\displaystyle (k+j+1)^{1/\alpha+1}},
\end{eqnarray*}
where $\alpha > 0$, $\beta > 0$, $r = 1, 2, \ldots$, and $C_{i}^{n}$ denotes the binomial coefficient $n!/(i!(n-i)!)$, see \cite{hosking1990moments}. So the first and the second $L$-moments are given by
$\mu_{1}^{L} = \beta \Gamma\left( {1/\alpha +1} \right)$
and
$\mu_{2}^{L} = \beta \Gamma \left( {1/\alpha +1} \right)\bigl({1-{\displaystyle 2^{-1/\alpha }}}\bigr)$, respectively.
The first two sample $L$-moments are:
\begin{eqnarray*}
m_{1}^{L} =\frac{1}{n}\sum\limits_{i=1}^n {X_{i:n}}=\overline{X},
\end{eqnarray*}
and
\begin{eqnarray*}
m_2^L =\frac{2}{n(n-1)}\sum\limits_{i=1}^n {(i-1) X_{i:n} - \overline {X}}.
\end{eqnarray*}
Now, equating $\mu_{1}^{L}$ and $\mu_{2}^{L}$ with $m_{1}^{L}$ and $m_{2}^{L}$, respectively, the $L$-moments of $\alpha$ and $\beta$ are obtained as:
\begin{eqnarray*}
\widehat{\alpha }_{LM}=-\frac{\displaystyle \ln (2)}{\displaystyle \ln \left( 1 - m_{2}^{L}/m_{1}^{L} \right)},
\end{eqnarray*}
and
\begin{eqnarray*}
\widehat {\beta }_{LM} =\frac {\displaystyle m_{1}^{L}}{\displaystyle \Gamma \left( {1/\widehat{\alpha }_{LM} +1} \right)}.
\end{eqnarray*}
\subsection{Method of logarithmic moment (MLM)}
The log-moment estimates of the shape and scale parameters of Weibull distribution with cdf (\ref{cdf}) are given by (see \cite{wayne1982applied}, \cite{norman1994continuous}, and \cite{dodson2006weibull})
\begin{eqnarray}
\widehat {\alpha}_{MLM} =\sqrt{\frac{\displaystyle \pi^2}{\displaystyle 6S^2}},
\label{26}
\end{eqnarray}
and
\begin{eqnarray}
\widehat {\beta}_{MLM} =\exp \big\{M_1 - \psi(1)/\widehat {\alpha }_{MLM}\big\},
\label{27}
\end{eqnarray}
where $S^2$ and $M_{1} $ are the sample variance and the mean of log-transformed data, respectively.
Also $\psi (1)=-0.5772156$.
It can be shown that (\ref{26}) and (\ref{27}) are both asymptotically
unbiased and consistent, see \cite{norman1994continuous}.
\subsection{Percentile method (PM)}
The quantile of a Weibull distribution with cdf (\ref{cdf}) is
\begin{eqnarray*}
x_{p} =\beta \left[ {-\ln (1-p)} \right]^{1/\alpha},
\end{eqnarray*}
where $0<p<1$, see \cite{norman1994continuous} and \cite{dodson2006weibull}. Using $p=1 - \exp (-1) \cong 0.632$, one can construct percentile-based estimators for $\alpha$ and $\beta$ as
\begin{eqnarray} \label{pmbeta}
\hat {\alpha }_{PM} =\left( {\frac{\displaystyle \ln [-\ln (1-p)]}{\displaystyle \ln \left(x_p\right) - \ln \left(x_{0.632}\right)}} \right),
\end{eqnarray}
and
\begin{eqnarray}\label{pmbeta}
\hat{\beta }_{PM} =x_{1 - \exp (-1)},
\end{eqnarray}
respectively, where $0<x_p <x_{0.632}$.
The suggested values for $p$ are 0.15 (see \cite{wang1995improved}) and 0.31, see (\cite{seki1996robust} and \cite{hassanein1971percentile}). Statistical tools show that
percentile-based estimators are, in general, asymptotically normal and unbiased, see \cite{wayne1982applied}.

\subsection{Method of moments (MM)}

Moment-based estimators of a given population are obtained by equating the population moments to their
sample counterparts and solving the resulting equations. The moment-based estimators for the Weibull distribution suffers from numerical computations, see \cite{cran1988moment}. Also, these estimators are not efficient. The $r$-th non-central moment for the Weibull distribution is (\cite{wayne1982applied}; \cite{norman1994continuous}; \cite{dodson2006weibull}):
\begin{eqnarray*}
\mu_r =\beta^r\Gamma \left(r/\alpha +1\right).
\end{eqnarray*}
Equating the mean and variance ($\mu_1$ and $\mu_2 - \mu_1^2$) with the sample
counterparts ($\overline {X}$ and $S^2$), the moment-based estimator of the shape parameter $\widehat{\alpha}_{MM}$, is root of the equation
\begin{eqnarray*}
\frac{\displaystyle \Gamma (1+2/\alpha )}{\displaystyle \Gamma^2(1+1/\alpha)}+\frac{\displaystyle S^2}{\displaystyle \overline{X}}-1=0,
\end{eqnarray*}
and the moment-based estimator of the scale parameter is
\begin{eqnarray*}
\widehat {\beta}_{MM}=\frac{\displaystyle \overline {X}}{\displaystyle \Gamma \left( {1/\widehat{\alpha}_{MM}+1} \right)}.
\end{eqnarray*}
\section{Performance comparisons}
This section has two parts. In the first part, we compare the performances of estimators introduced in Section 2 and 3 through simulation. Second part devoted to an illustration in which all estimators are applied to a set of real data.
\subsection{Simulation study}
Here, we perform a Monte Carlo simulation to compare the performance of the $U$-statistic, MLE, WMLE, GLS1, GLS2, WLS, LM, MLM, PM, and MM. For this aim, we compare the bias and root of mean squared error (RMSE). \par For computing the bias we adopt small sizes of sample  including 5, 10, 30, and two levels for shape and scale parameters as: (0.5, 0.5), (2.5, 0.5), (0.5, 2.5), and (2.5, 2.5). The results after computing the bias are given in Tables \ref{tab1}-\ref{tab2}. Also the bias of $U$-statistic, MLE, GLS1, GLS2, WLS, and LM are given for large sizes of sample including 1000 and 4000. The corresponding results are given in Tables \ref{tab3}-\ref{tab4} for shape and scale parameters, respectively. \par
\begin{figure}
\includegraphics[width=50mm,height=50mm]{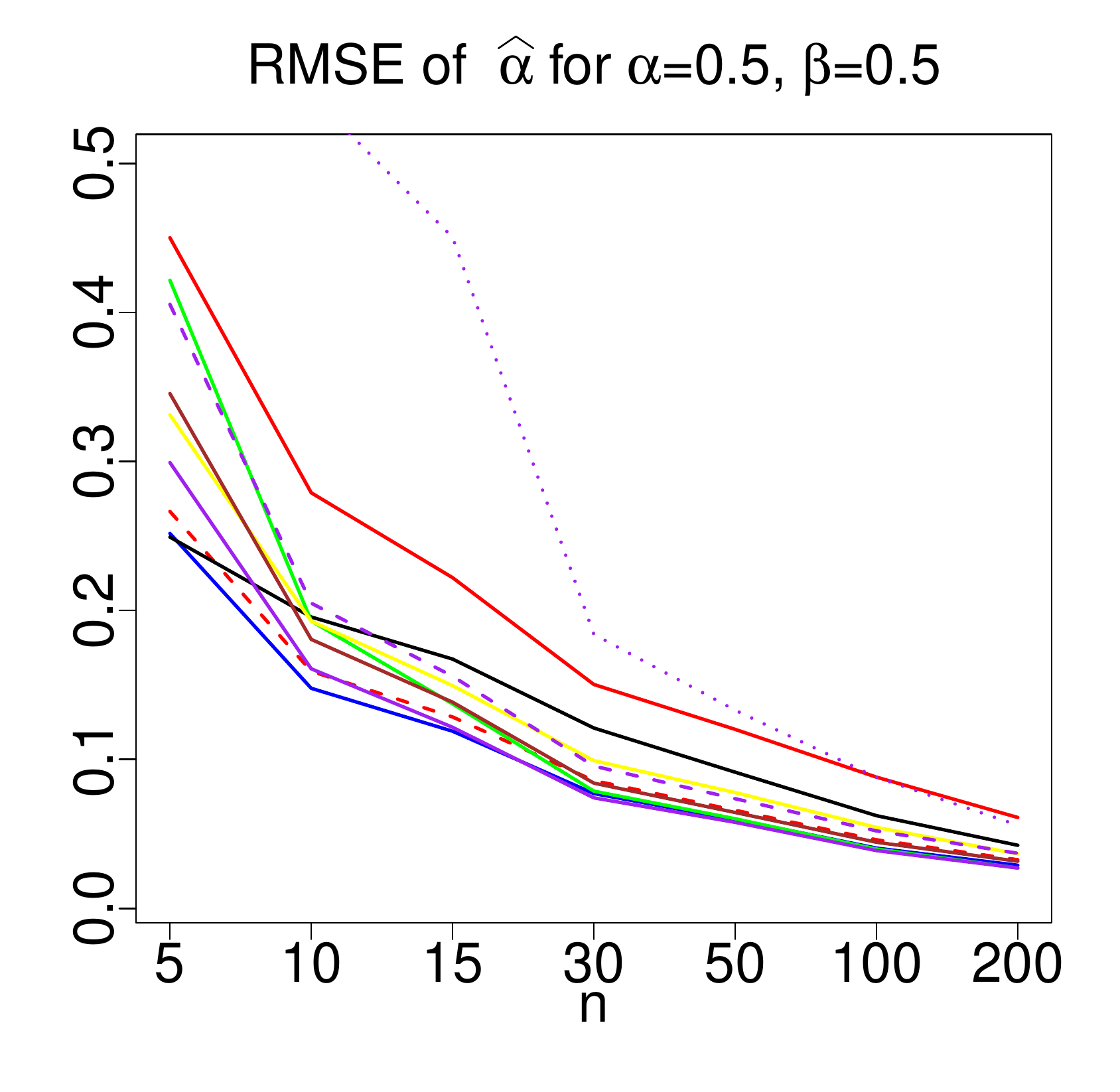}
\includegraphics[width=50mm,height=50mm]{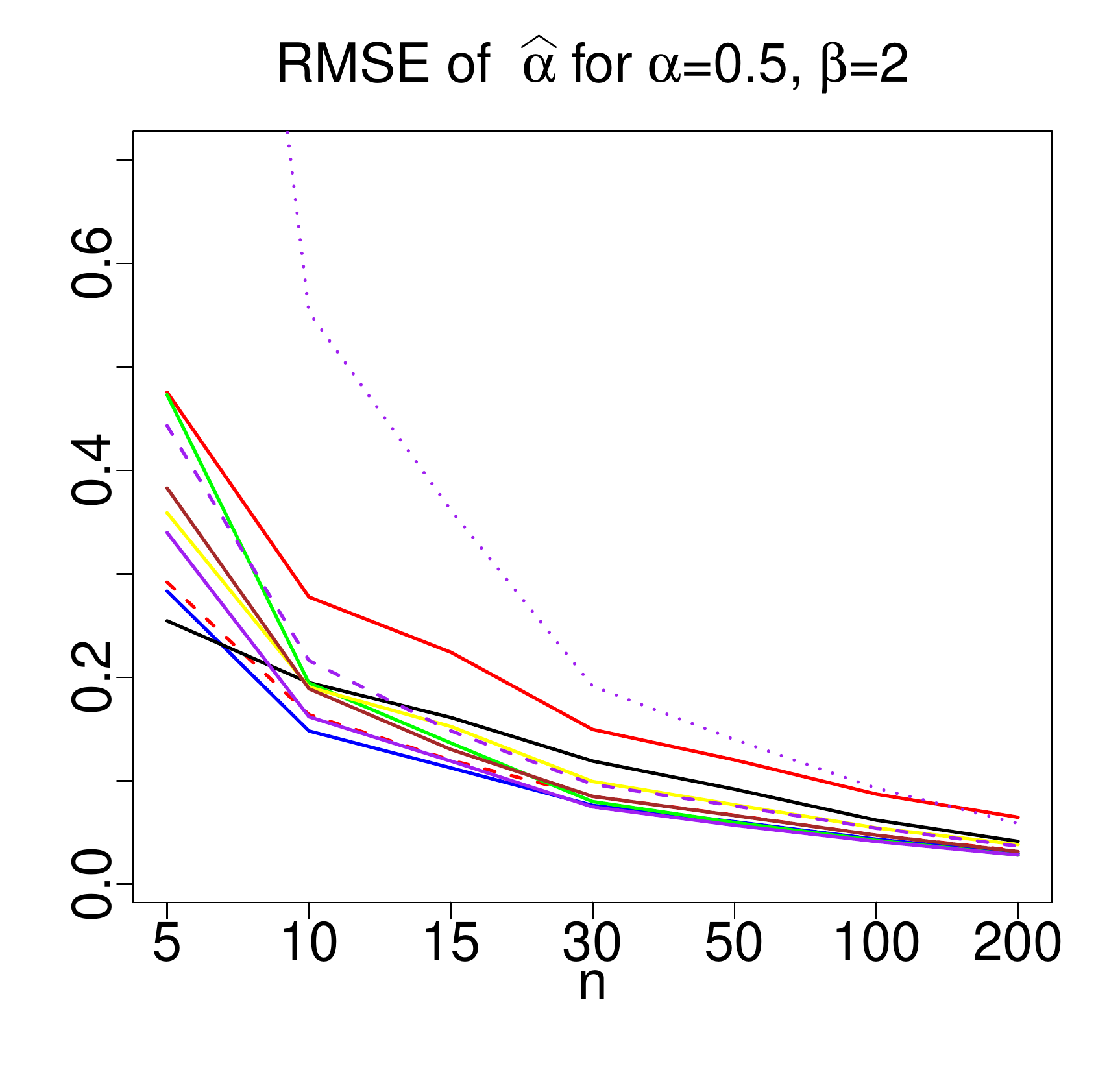}
\includegraphics[width=50mm,height=50mm]{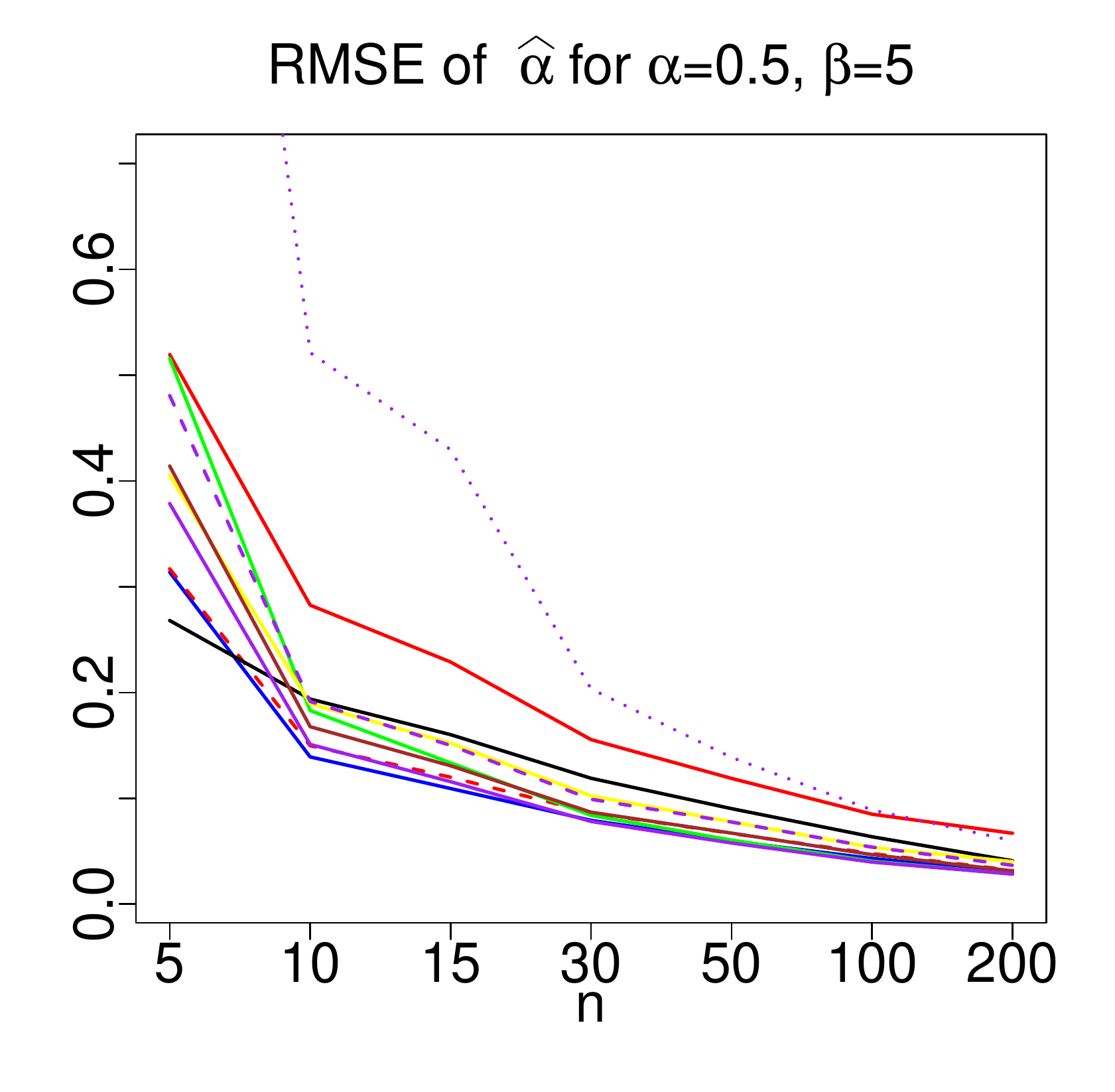}\\
\includegraphics[width=50mm,height=50mm]{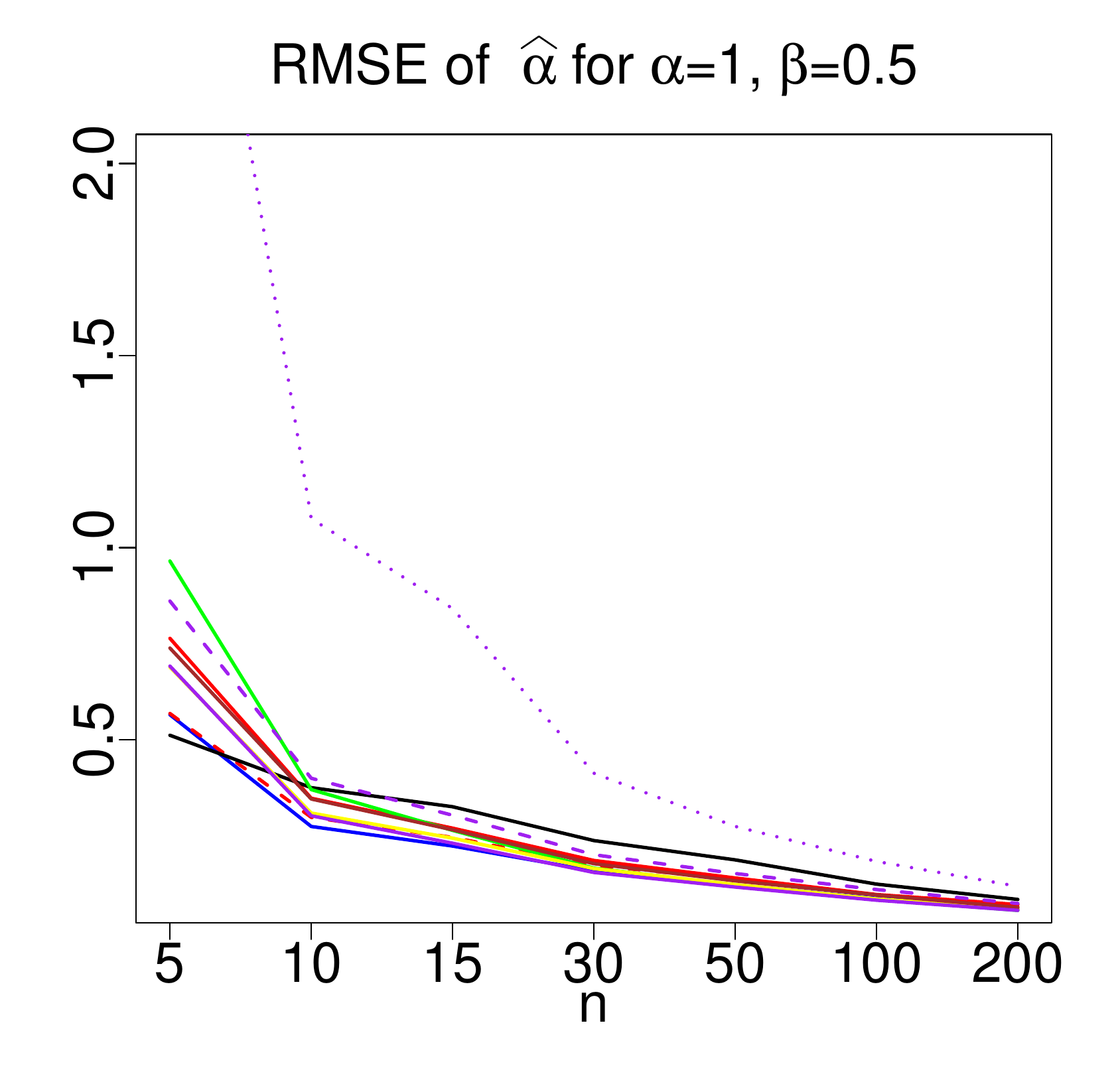}
\includegraphics[width=50mm,height=50mm]{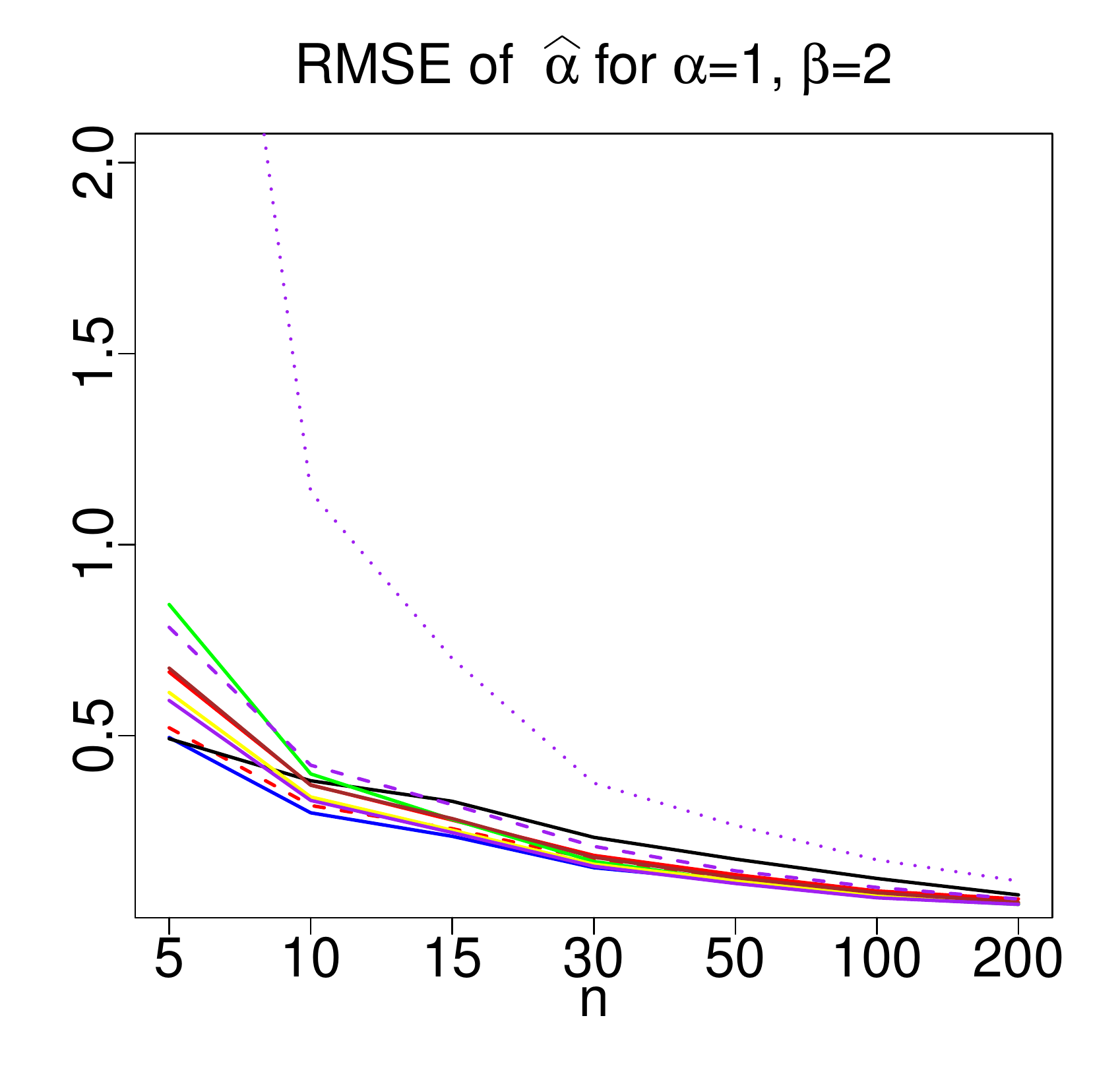}
\includegraphics[width=50mm,height=50mm]{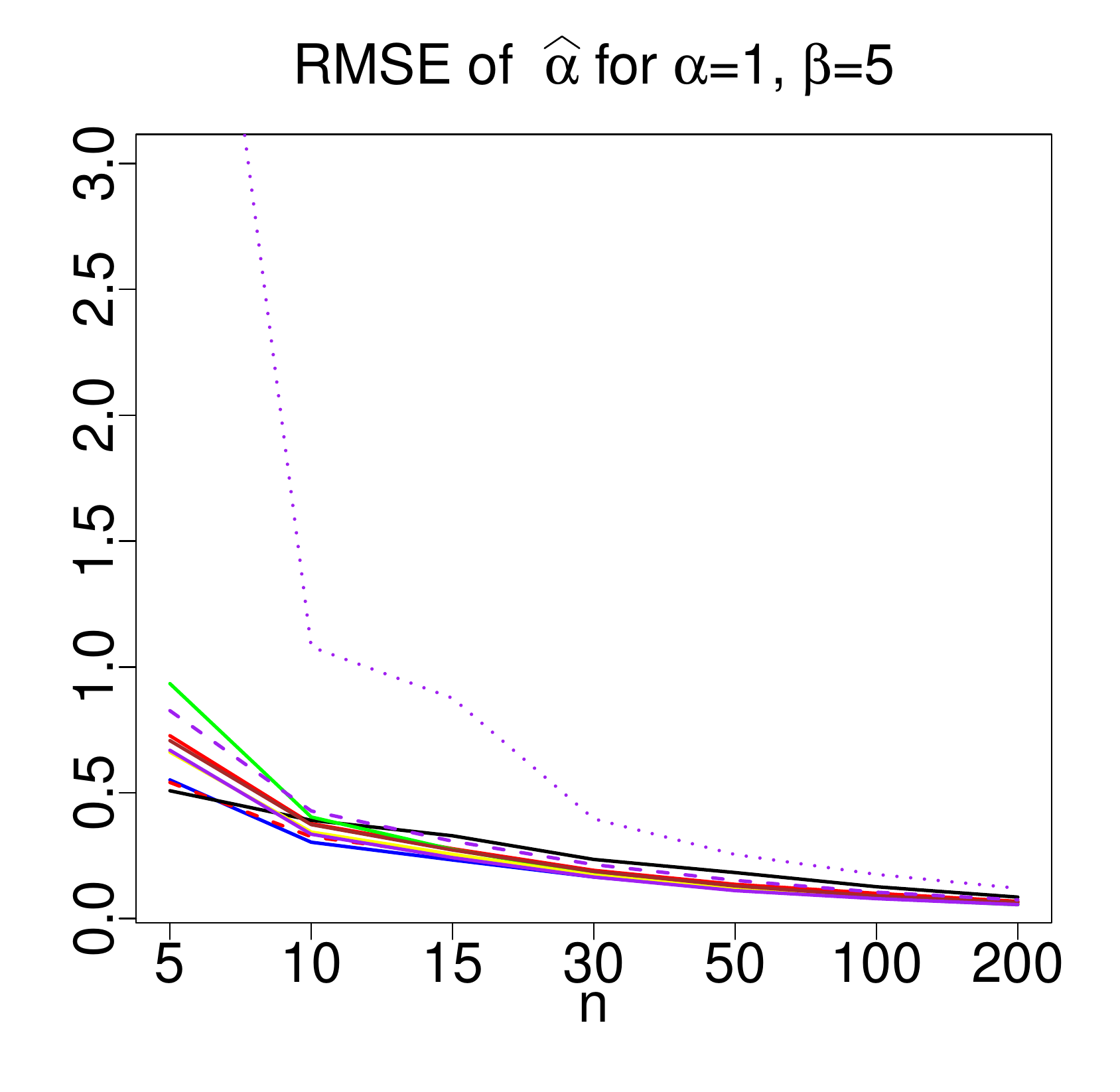}\\
\includegraphics[width=50mm,height=50mm]{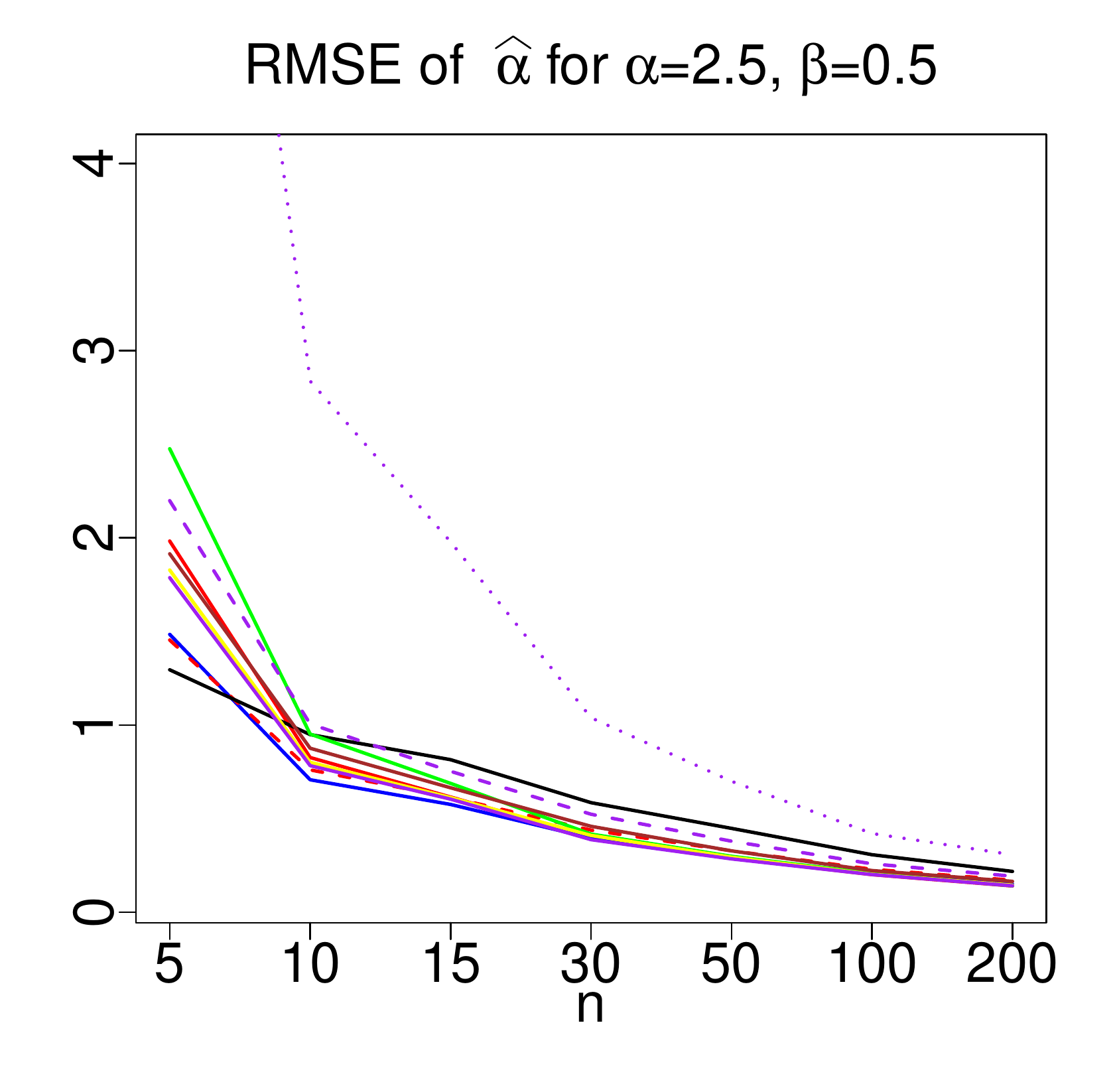}
\includegraphics[width=50mm,height=50mm]{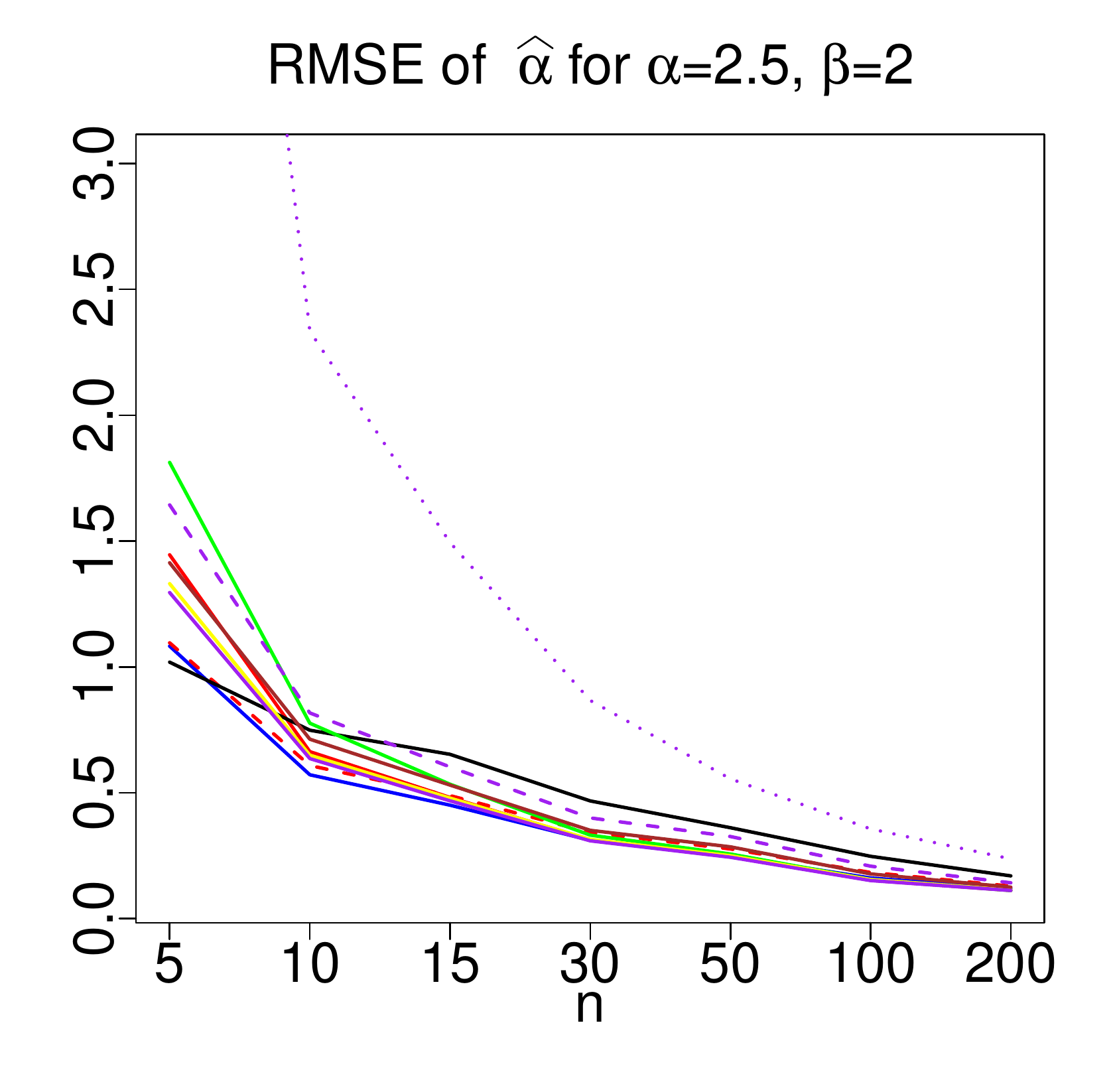}
\includegraphics[width=50mm,height=50mm]{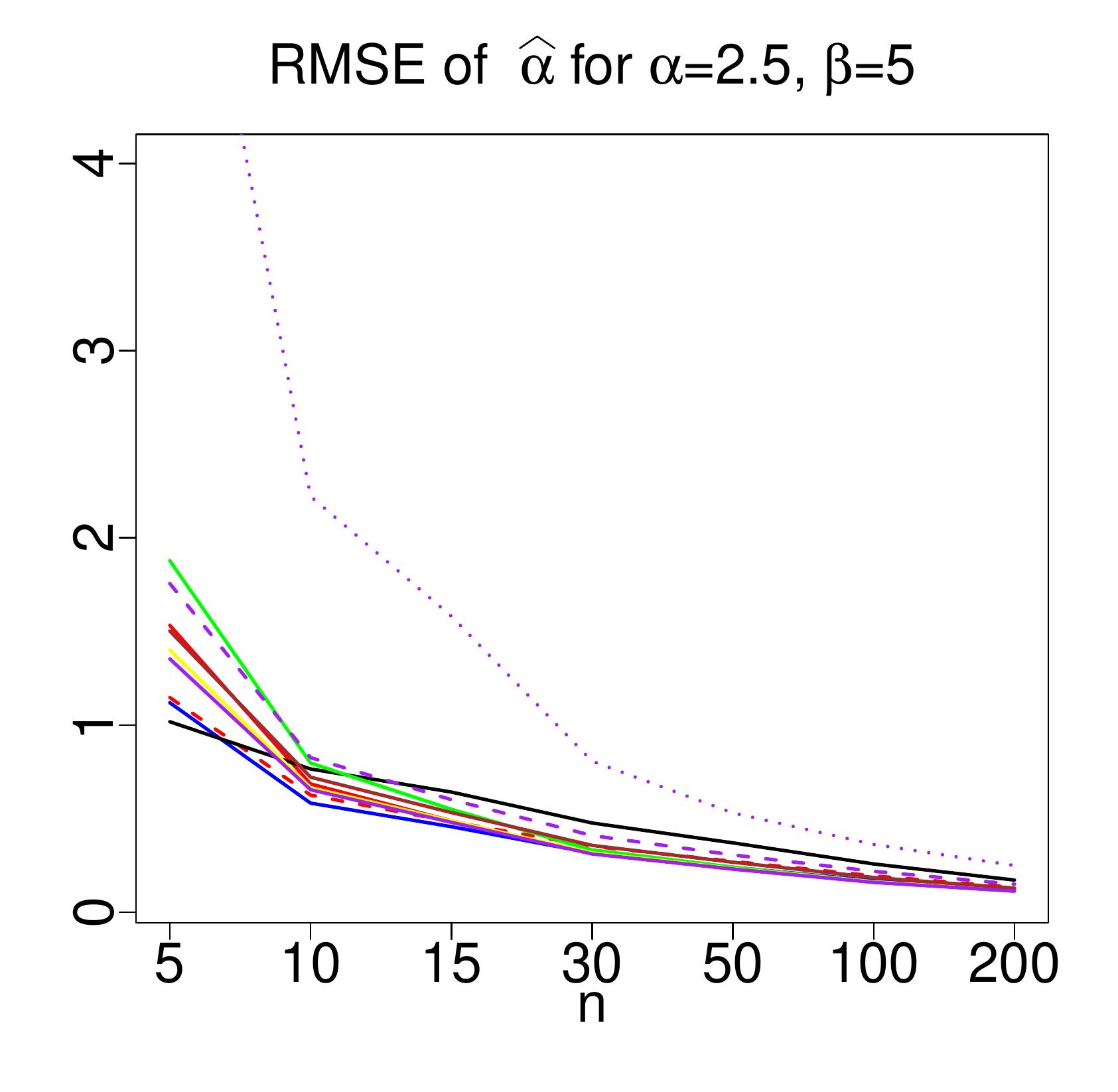}\\
\caption{\tiny{RMSE of the shape parameter estimator, $\hat{\alpha}$ for different levels of $\alpha$ and $\beta$ under small sample size scenario, i.e., $n=$5, 10, 15, 30, 50, 100, and 200. The used color scheme are: the black solid curve for the GLS2, blue solid curve for the WLS, brown solid curve for the $U$-statistic, green solid curve for the MLE, solid red curve for the MM, dashed red curve for the GLS1, purple solid curve for the WMLE, dotted purple curve for the PM, dashed purple curve for the MLM, and yellow solid curve for LM.}}
\label{fig1}
\end{figure}

\begin{figure}
\includegraphics[width=50mm,height=50mm]{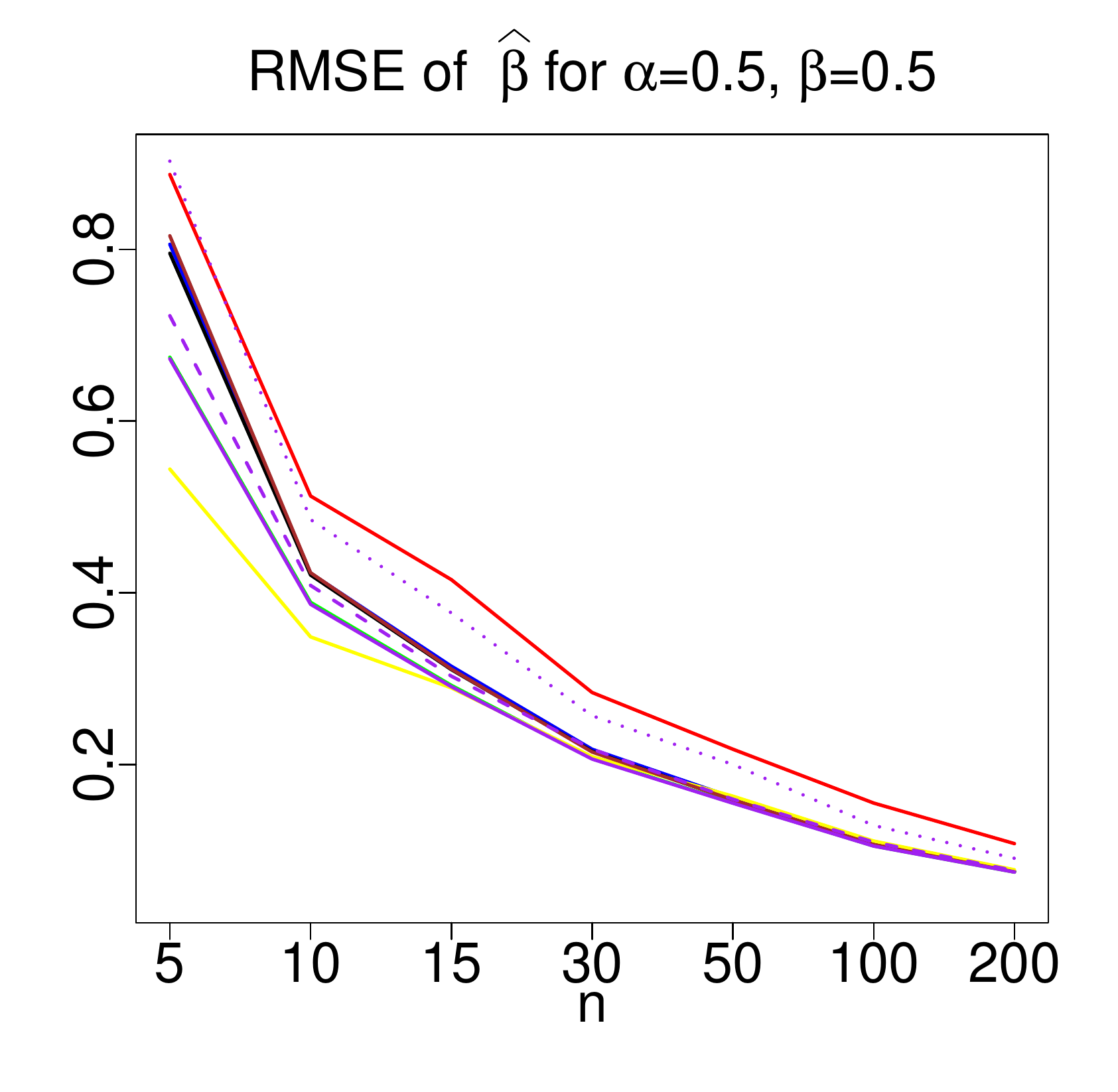}
\includegraphics[width=50mm,height=50mm]{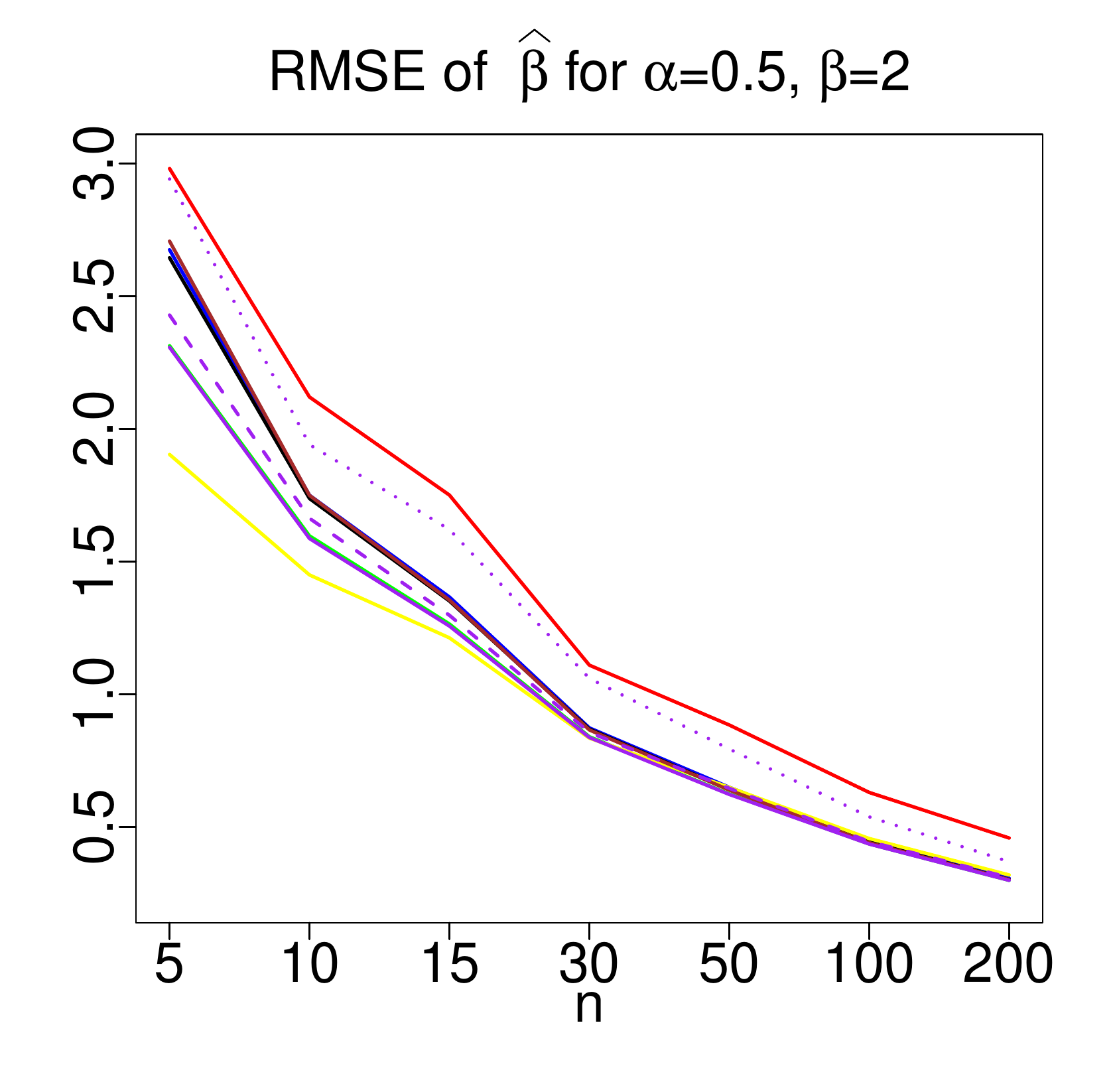}
\includegraphics[width=50mm,height=50mm]{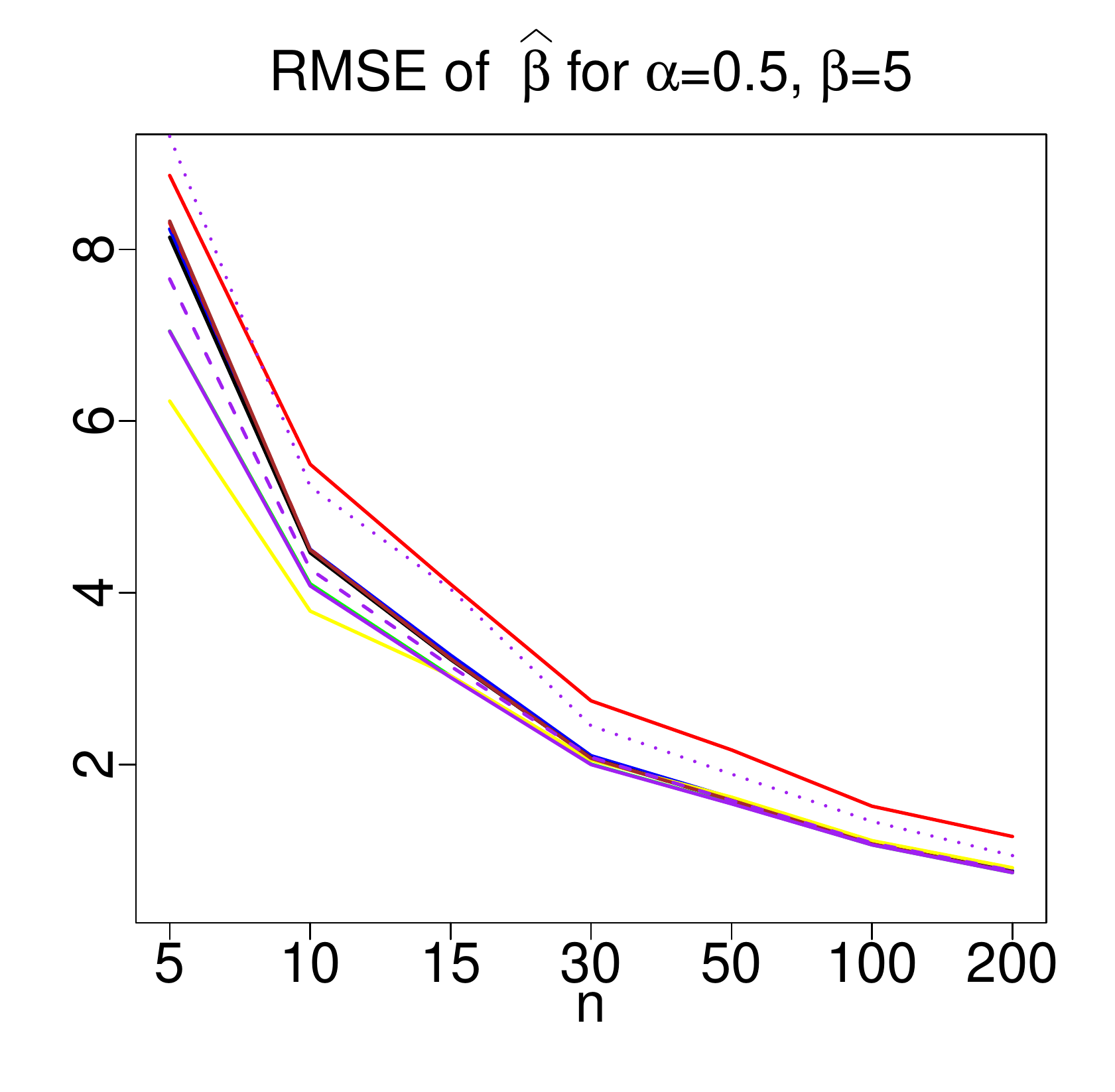}\\
\includegraphics[width=50mm,height=50mm]{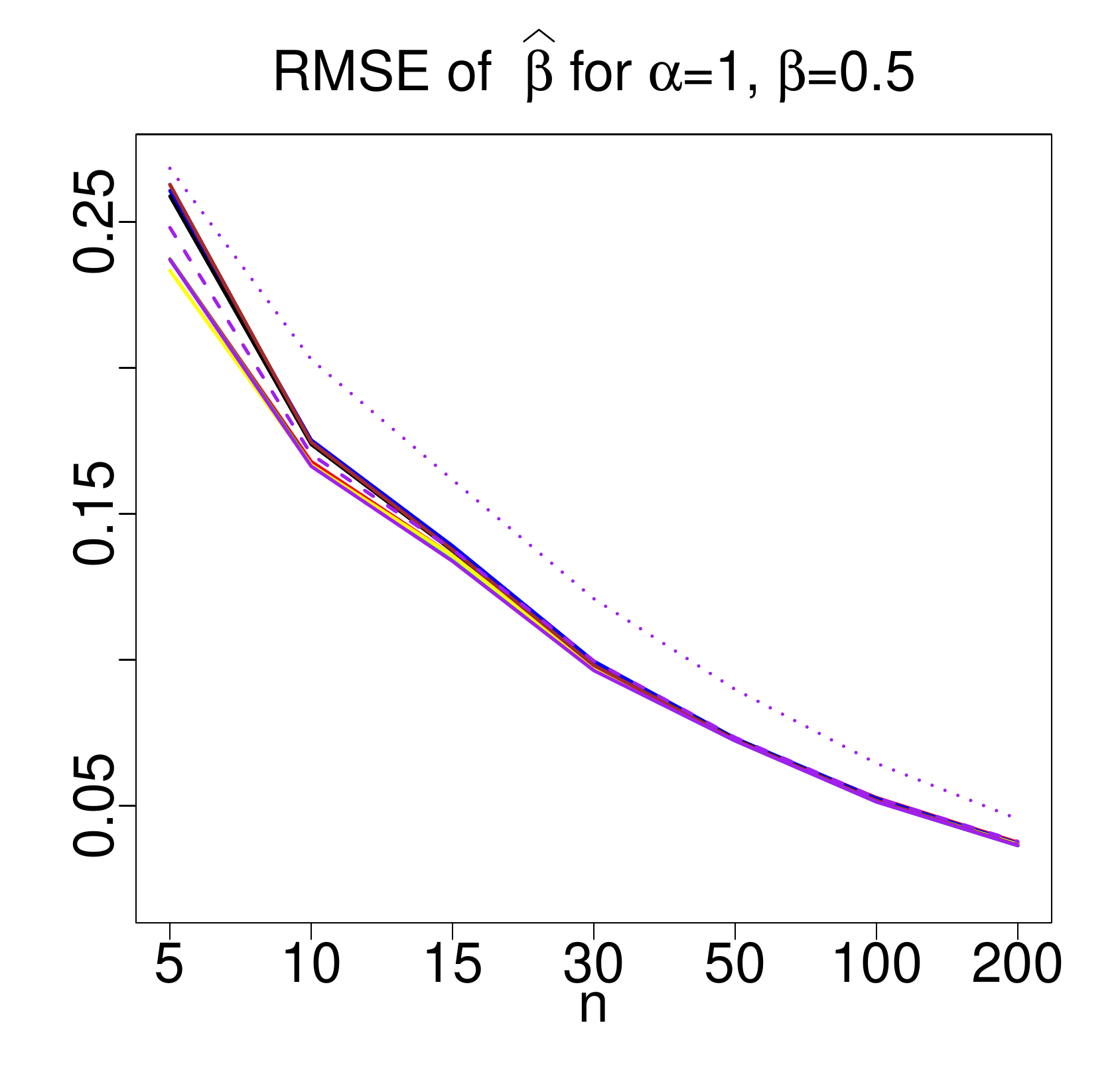}
\includegraphics[width=50mm,height=50mm]{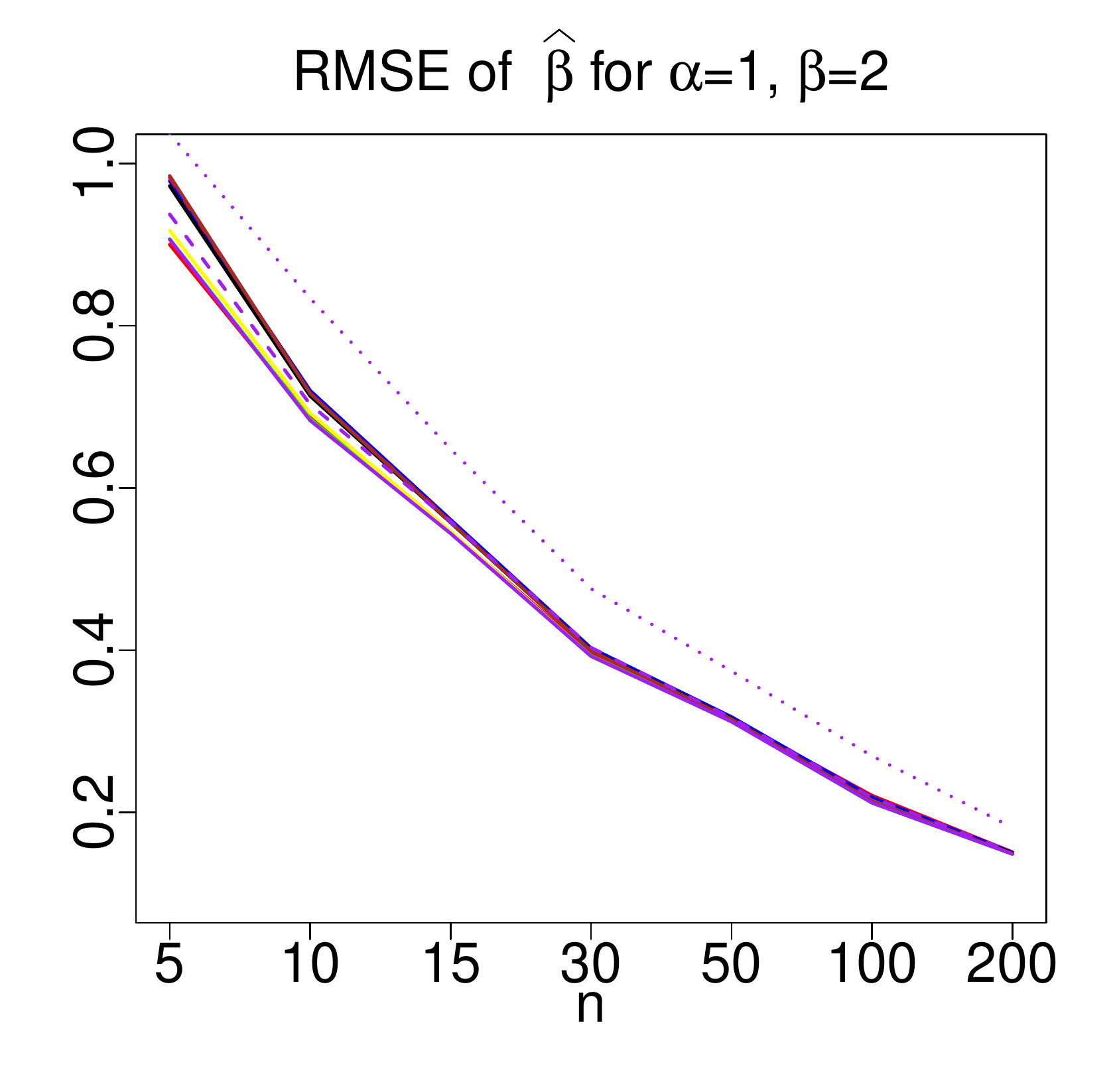}
\includegraphics[width=50mm,height=50mm]{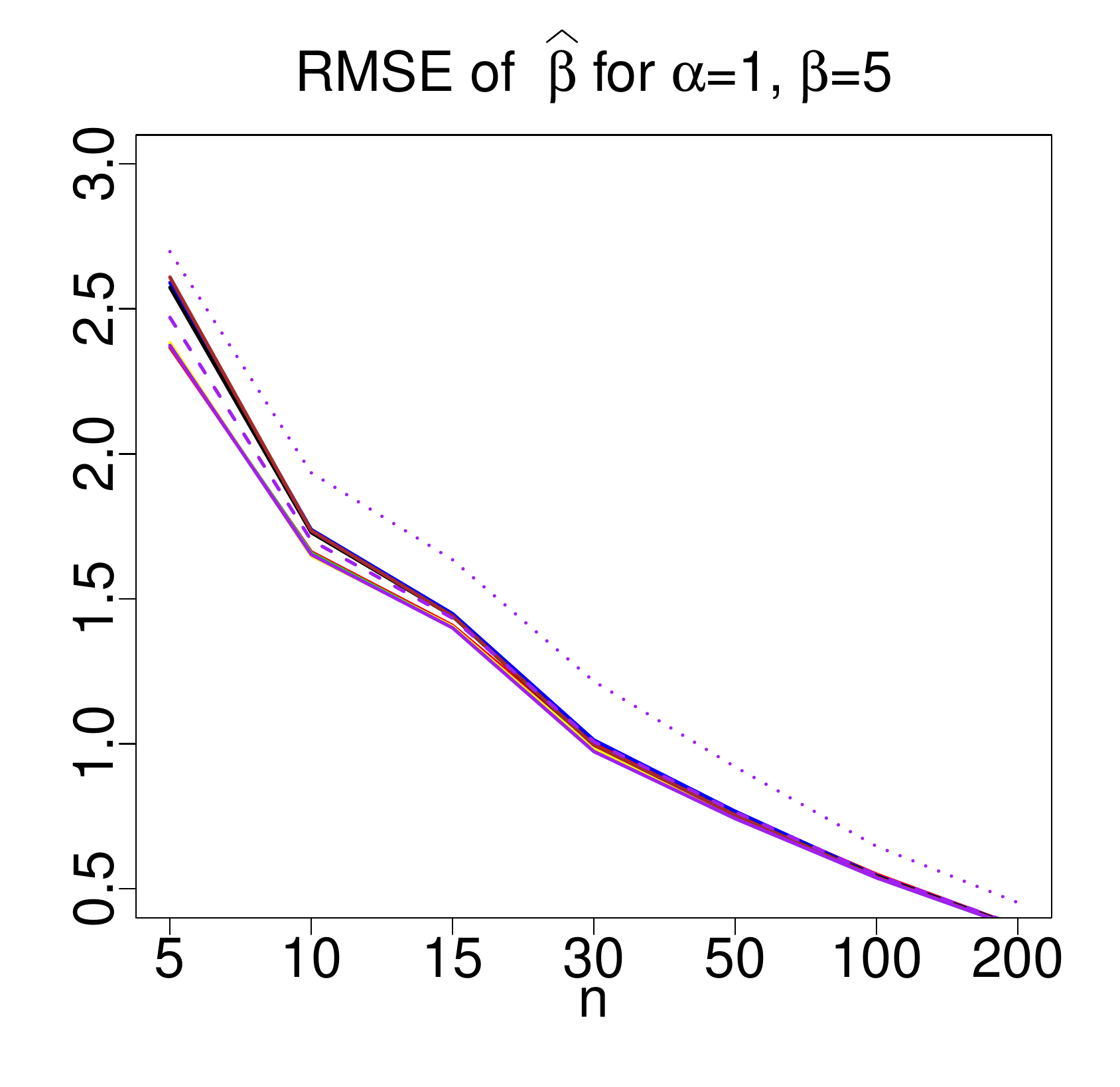}\\
\includegraphics[width=50mm,height=50mm]{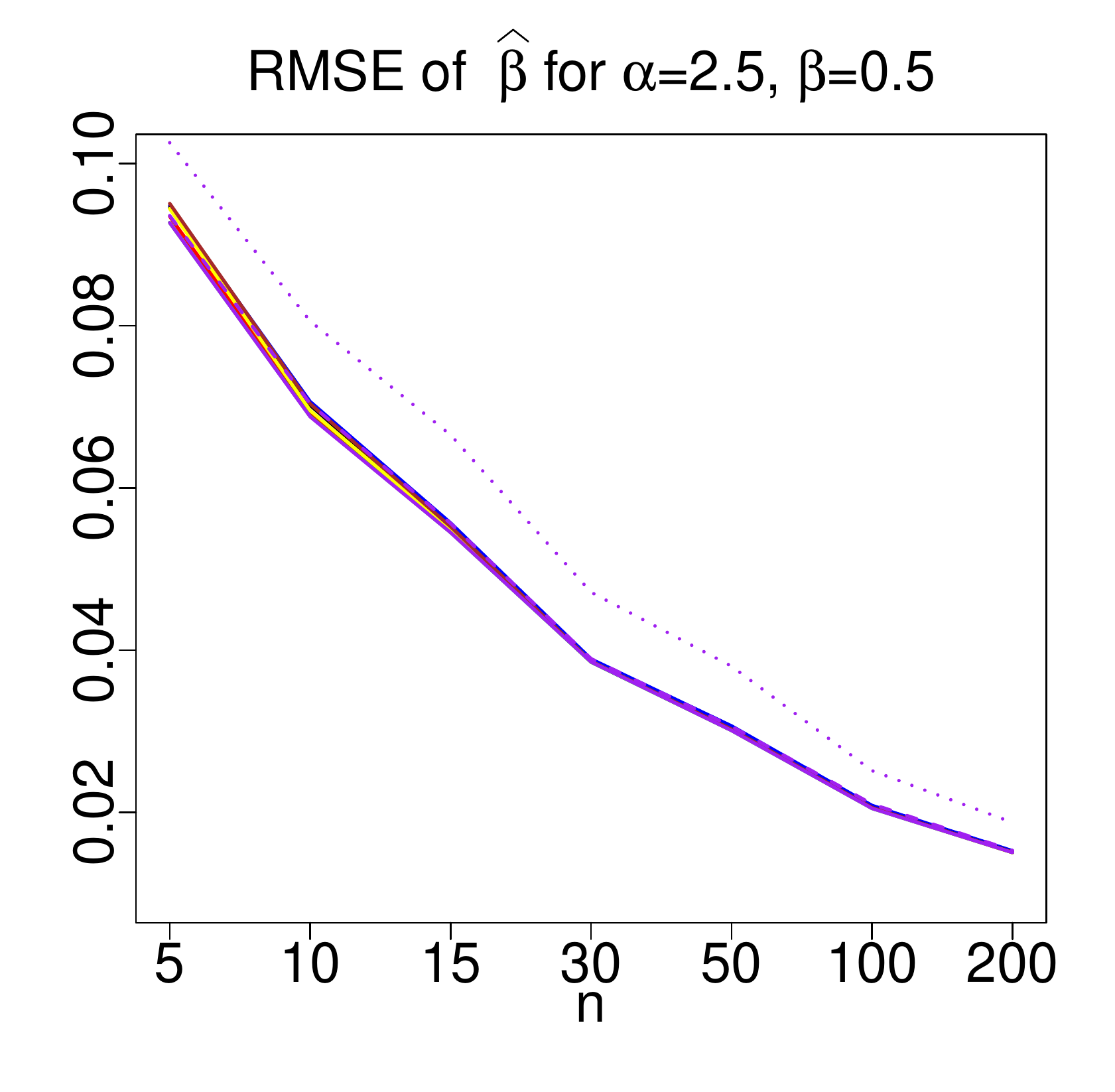}
\includegraphics[width=50mm,height=50mm]{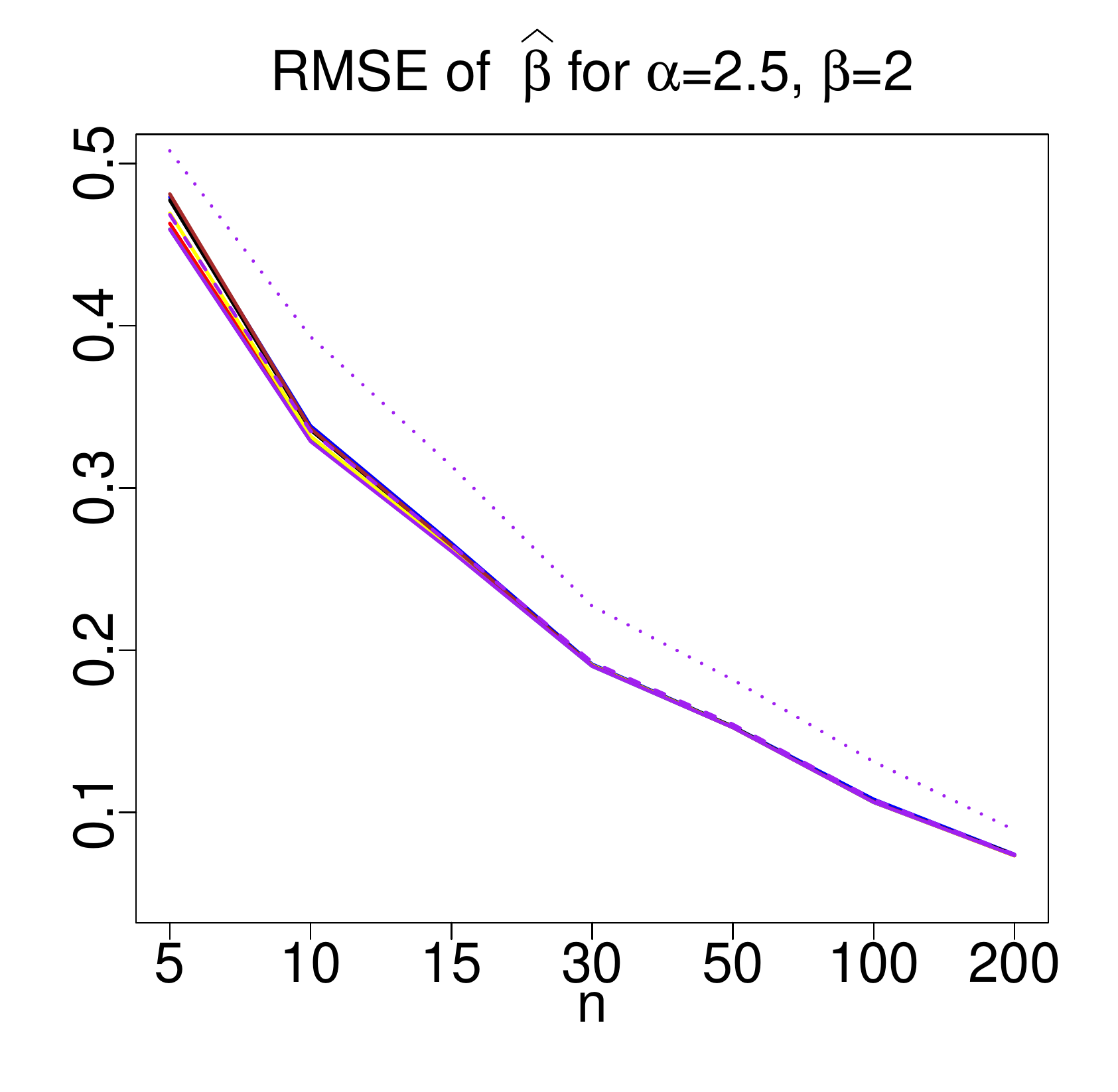}
\includegraphics[width=50mm,height=50mm]{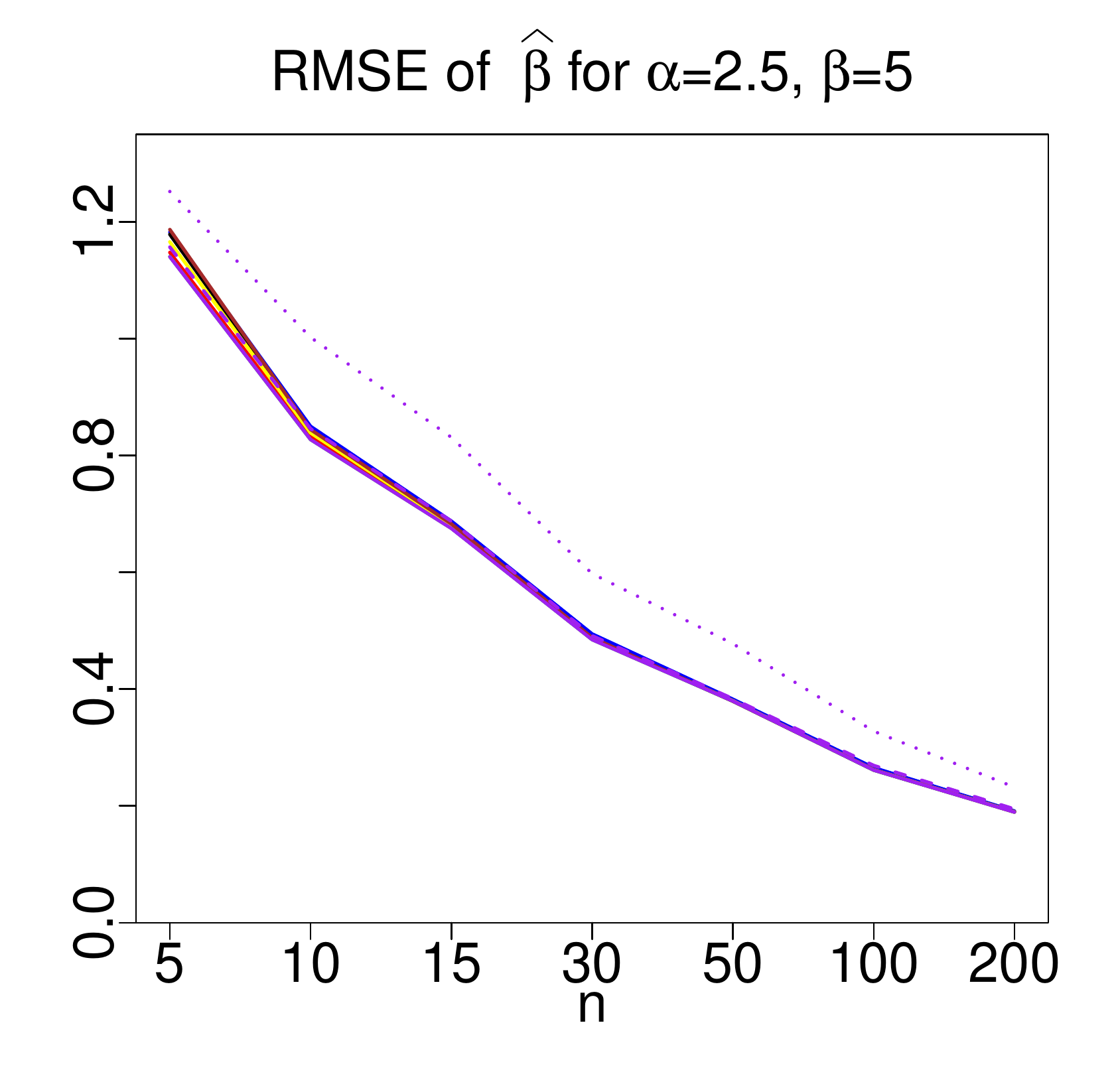}\\
\caption{\tiny{RMSE of the scale parameter estimator, $\hat{\beta}$ for different levels of $\alpha$ and $\beta$ under small sample size scenario, i.e., $n=$5, 10, 15, 30, 50, 100, and 200. The used color scheme are: the black solid curve for the GLS2, blue solid curve for the WLS, brown solid curve for the $U$-statistic, green solid curve for the MLE, solid red curve for the MM, dashed red curve for the GLS1, purple solid curve for the WMLE, dotted purple curve for the PM, dashed purple curve for the MLM, and yellow solid curve for LM.}}
\label{fig2}
\end{figure}

 For computing the RMSE, we choose the sample sizes as: 5, 10, 15, 30, 50, 100, and 200. Comparisons are performed for different levels of the shape ($\alpha$=0.5, 1, and 2.5) and the scale ($\beta$=0.5, 2, and 5) parameters. We used a 7-color scheme to distinguish between competitors through Figures \ref{fig1}-\ref{fig2} as follows. The brown for the $U$-statistic, green for the MLE, purple for the WMLE, dashed red for the GLS1, black for the GLS2, blue for the WLS, dashed purple for the MLM, dotted purple for the PM, solid red for the MM, and yellow solid curve for LM. The results for computing the RMSE are given in Figures \ref{fig1}-\ref{fig2}.

%In the second scenario, the comparisons are made between bias and RMSE of GLS2 (black line), WLS (red line), $U$-statistic (brown line), MLE (green line), GLS1 (blue line), and LM (yellow line) which are shown in Figures \ref{fig5}-\ref{fig8}. The WML, MM, MLM, and PM estimators are eliminated by competitions. This is because MLE and WMLE have the same performances and MM, MLM, and PM have the weak performances compared with the other estimators. The performance of these six estimators are compared for samples of sizes 1000, 2000, and 4000 for $N$=500 iterations. In both scenarios, the comparisons are based on the bias and the root mean square error (RMSE).

\subsubsection{Comparison results for the bias}
According to the bias of the shape parameter estimator $\hat{\alpha}$ for small sizes 5, 10, and 30, the following conclusions can be made from Table \ref{tab1}.
\begin{enumerate}
\item GLS2 , WLS, and WMLE give the best performances for $n=5$, $n=10$, and $n=30$, respectively.
\item
WMLE shows the best performance next to the WLS and GLS1.
\item
PM gives the worst performance.
\item
When $\alpha$ is small (say $\alpha = 0.5$), the MM gives the worst performance next to the GLS2.
\item
When $\alpha$ is large (say $\alpha = 2.5$) and $n\geq 15$, the GLS2 gives the worst performance.
\item
WMLE outperforms the LM.
\item
WMLE and $U$-statistic outperform the MLE.
\item
$U$-statistic shows better performance than the MLE, MLM, MM, and PM.
\end{enumerate}
The following observations can be made from Table \ref{tab2} for bias of the scale parameter estimator $\hat{\beta}$ for small sizes 5, 10, and 30.
\begin{enumerate}
\item WMLE and MLE give almost the same performances.
\item
When $\alpha$ is small (say $\alpha = 0.5$), the MM gives the worst performance.
\item
When $\alpha$ is small, the LM gives the best performance.
\item
The GLS2, GLS1, and WLS show the same performances.
\item
MLM outperforms GLS2, GLS1, and WLS.
\item
When $\alpha$ is large (say $\alpha=2.5$), the PM shows the worst performance.
%When $\alpha$ is not large (say $\alpha \leq 1.5$), the MLM outperforms GLS2, GLS1, and WLS.
\item
The GLS2, GLS1, and WLS outperforms the $U$-statistic for $n=5, 10$.
\end{enumerate}
The following observations can be made from Tables \ref{tab3}-\ref{tab4} for bias of the shape parameter estimator $\hat{\beta}$ for large sizes 1000 and 4000.
\begin{enumerate}
\item $U$-statistic gives the best performance for estimating the shape and scale parameters.
\item
GLS1 shows the worst performance for estimating the shape and scale parameters.
\end{enumerate}
We note that for bias analysis when sample sizes are large, the MLM, PM, and MM have been eliminated by competitions since the show weak performances. Also, since MLE and WMLE show the same performances, the WMLE has been removed by competitions.
\subsubsection{Comparison results for RMSE}
The following observations can be made from Figure \ref{fig1} for RMSE of the shape parameter estimator $\hat{\alpha}$.
\begin{enumerate}
\item
The PM gives the worst performance.
\item
When $n=5$ the GLS2 gives the best performance.
\item
When $n=5$ the GLS2 gives the best performance.
\item
The WGLS gives the best performance next to the GLS1.
\item
The WMLE outperforms the LM and $U$-statistic.
\item
The MLM shows better performance than the MLE for sample size (say $n \leq 10$).
\end{enumerate}
The following observations can be made from Figure \ref{fig2} for RMSE of the scale parameter estimator $\hat{\beta}$.
\begin{enumerate}
\item
The PM gives the worst performance.
\item
When $\alpha$ is small (say $\alpha \leq 0.5$), the MM gives the worst performance.
\item
When $\alpha$ is not small (say $\alpha \geq 1$), the PM gives the worst performance.
\item
When $\alpha$ is small (say $\alpha \leq 0.5$) and $n\leq 15$, the LM gives the best performance.
\end{enumerate}
\subsection{Real data illustration}
Here, we apply all reviewed methods introduced in Sections 2 and 3 to a set of real data involving by lifetimes in years reported by \cite[p. 17]{breaking}. Data are shown in Table \ref{tab5}. To implement these techniques, programs have been written in \verb|R| environment, see \cite{team2014r}. In order to compare the performance of estimators presented in the Section 2 and 3, we employed the Kolmogorov-Smirnov (KS) and Cramer-Von Mises (CVM) distances which are given by
\begin{equation*}
\label{KS}
\text{KS}=\max\limits_{\boldsymbol{1 \leq i \leq n}} \max\left\{\frac{i}{n}-F_X\bigl(x_{(i)}\bigr), F_X\bigl(x_{(i)}\bigr)-\frac{i-1}{n}\right\},
\end{equation*}
and
\begin{equation*}
\label{vm}
\text{CVM}=\frac{1}{12 n}+\sum\limits_{i=1}^n\biggl[ \frac{2i-1}{2n} -F_X\bigl(x_{(i)}\bigr)\biggr]^2,
\end{equation*}
where $n$ is the sample size, $x_{(i)}$; for $i=1\dots, n$, is the $i$-th ordered observed value and $F_X(.)$ is the distribution function of two-parameter Weibull distribution defined in (\ref{cdf}).
The following observations can be made from Table \ref{tab6}.
\begin{enumerate}
\item The WLS shows the best performance in the sense of both criteria KS and CVM.
\item The MLM shows the best performance in the sense of CVM criterion next to the WLS.
\item The PM shows the best performance in the sense of KS criterion next to the WLS.
\end{enumerate}

\begin{table}
%\tiny
\caption{Bias of shape parameter estimators for samples of small size.}
\begin{center}
\begin{tabular}{lccccccc}
\cline{1-5}
&\multicolumn{4}{c}{n=5}\\ \cline{1-5}
&\multicolumn{4}{c}{parameters level}\\ \cline{1-5}
Method&($\alpha=0.5$, $\beta=0.5$) &($\alpha=0.5$, $\beta=2.5$)& ($\alpha=2.5$, $\beta=0.5$)& ($\alpha=2.5$, $\beta=2.5$)\\ \cline{1-5}
$U$-Statistic&0.346& 0.383& 1.890& 1.415\\
MLE &0.422& 0.473& 2.453& 1.813\\
WMLE &0.299& 0.340& 1.764& 1.309\\
GLS1 &0.266& 0.292& 1.420& 1.106\\
GLS2 &0.249& 0.255& 1.261& 1.025\\
WLS &0.252& 0.283& 1.449& 1.082\\
LM &0.331& 0.359& 1.798& 1.353\\
MLM &0.405& 0.443& 2.203& 1.645\\
PM &1.199& 1.699& 8.597& 7.132\\
MM &0.450& 0.476& 1.971& 1.452\\
\cline{1-5} \\
\cline{1-5}
&\multicolumn{4}{c}{n=10}\\ \cline{1-5}
&\multicolumn{4}{c}{parameters level}\\ \cline{1-5}
Method&($\alpha=0.5$, $\beta=0.5$) &($\alpha=0.5$, $\beta=2.5$)& ($\alpha=2.5$, $\beta=0.5$)& ($\alpha=2.5$, $\beta=2.5$)\\ \cline{1-5}
$U$-Statistic&0.181& 0.189& 0.865& 0.713\\
MLE &0.193& 0.195& 0.914& 0.776\\
WMLE &0.161& 0.162& 0.756& 0.636\\
GLS1 &0.159& 0.164& 0.765& 0.608\\
GLS2 &0.196& 0.195& 0.966& 0.749\\
WLS &0.148& 0.146& 0.697& 0.572\\
LM &0.193& 0.191& 0.781& 0.649\\
MLM &0.205& 0.216& 0.999& 0.817\\
PM &0.546& 0.553& 2.491& 2.334\\
MM &0.279& 0.278& 0.806& 0.664\\
\cline{1-5}\\\cline{1-5}
&\multicolumn{4}{c}{n=30}\\ \cline{1-5}
&\multicolumn{4}{c}{parameters level}\\ \cline{1-5}
Method&($\alpha=0.5$, $\beta=0.5$) &($\alpha=0.5$, $\beta=2.5$)& ($\alpha=2.5$, $\beta=0.5$)& ($\alpha=2.5$, $\beta=2.5$)\\ \cline{1-5}
$U$-Statistic &0.071& 0.074 & 0.383& 0.301\\
MLE &0.078& 0.079 & 0.410& 0.332\\
WMLE &0.074& 0.073 & 0.379& 0.309\\
GLS1 &0.085& 0.084 & 0.425& 0.342\\
GLS2 &0.121& 0.119 & 0.588& 0.468\\
WLS &0.077& 0.076 & 0.385& 0.311\\
LM &0.099& 0.110 & 0.393& 0.315\\
MLM &0.095& 0.096 & 0.493& 0.400\\
PM &0.184& 0.191& 1.045& 0.867\\
MM &0.150& 0.149 & 0.386& 0.313\\
\cline{1-5}
\end{tabular}
\end{center}
\label{tab1}
\end{table}

\begin{table}
%\tiny
\caption{Bias of scale parameter estimators for samples of small size.}
\begin{center}
\begin{tabular}{lccccccc}
\cline{1-5}
&\multicolumn{4}{c}{n=5}\\ \cline{1-5}
&\multicolumn{4}{c}{parameters level}\\ \cline{1-5}
Method&($\alpha=0.5$, $\beta=0.5$) &($\alpha=0.5$, $\beta=2.5$)& ($\alpha=2.5$, $\beta=0.5$)& ($\alpha=2.5$, $\beta=2.5$)\\ \cline{1-5}
$U$-Statistic& 0.816& 2.717 & 0.094& 0.481\\
MLE & 0.674& 2.318 & 0.091& 0.460\\
WMLE & 0.673& 2.315 & 0.091& 0.459\\
GLS1 & 0.809& 2.686 & 0.094& 0.481\\
GLS2 & 0.795& 2.641 & 0.093& 0.477\\
WLS & 0.806& 2.670 & 0.094& 0.479\\
LM & 0.544& 1.909 & 0.093& 0.469\\
MLM & 0.723& 2.431 & 0.092& 0.468\\
PM & 0.903& 2.943 & 0.099& 0.508\\
MM & 0.927& 2.983 & 0.092& 0.463\\
\cline{1-5} \\
\cline{1-5}
&\multicolumn{4}{c}{n=10}\\ \cline{1-5}
&\multicolumn{4}{c}{parameters level}\\ \cline{1-5}
Method&($\alpha=0.5$, $\beta=0.5$) &($\alpha=0.5$, $\beta=2.5$)& ($\alpha=2.5$, $\beta=0.5$)& ($\alpha=2.5$, $\beta=2.5$)\\ \cline{1-5}
$U$-Statistic& 0.424 & 1.752& 0.067& 0.337\\
MLE & 0.389 & 1.608& 0.066& 0.329\\
WMLE & 0.387 & 1.596& 0.066& 0.329\\
GLS1 & 0.385 & 1.652& 0.067& 0.337\\
GLS2 & 0.421 & 1.744& 0.067& 0.336\\
WLS & 0.423 & 1.752& 0.067& 0.338\\
LM & 0.349 & 1.455& 0.067& 0.332\\
MLM & 0.409 & 1.663& 0.067& 0.335\\
PM & 0.485 & 1.948& 0.077& 0.393\\
MM & 0.513 & 2.129& 0.066& 0.331\\
\cline{1-5} \\ \cline{1-5}
&\multicolumn{4}{c}{n=30}\\ \cline{1-5}
&\multicolumn{4}{c}{parameters level}\\ \cline{1-5}
Method&($\alpha=0.5$, $\beta=0.5$) &($\alpha=0.5$, $\beta=2.5$)& ($\alpha=2.5$, $\beta=0.5$)& ($\alpha=2.5$, $\beta=2.5$)\\ \cline{1-5}
$U$-Statistic & 0.215& 0.866& 0.038& 0.186\\
MLE & 0.207& 0.841& 0.037& 0.187\\
WMLE & 0.207& 0.842& 0.037& 0.188\\
GLS1 & 0.214& 0.869& 0.038& 0.191\\
GLS2 & 0.215& 0.869& 0.038& 0.191\\
WLS & 0.215& 0.874& 0.038& 0.191\\
LM & 0.201& 0.830& 0.037& 0.188\\
MLM & 0.218& 0.857& 0.038& 0.195\\
PM & 0.257& 1.060& 0.045& 0.227\\
MM & 0.284& 1.110& 0.037& 0.202\\
\cline{1-5}
\end{tabular}
\end{center}
\label{tab2}
\end{table}

\begin{table}
%\tiny
\caption{Bias of shape parameter estimators for samples of large size.}
\begin{center}
\begin{tabular}{lccccccc}
\cline{1-5}
&\multicolumn{4}{c}{n=1000}\\ \cline{1-5}
&\multicolumn{4}{c}{parameters level}\\ \cline{1-5}
Method&($\alpha=0.5$, $\beta=0.5$) &($\alpha=0.5$, $\beta=2.5$)& ($\alpha=2.5$, $\beta=0.5$)& ($\alpha=2.5$, $\beta=2.5$)\\ \cline{1-5}
GLS1 & 0.005193& 0.003696& 0.018966& 0.008460\\
WLS & 0.005025 & 0.003351& 0.017895& 0.006642\\
GLS2 & -0.005167& -0.002230& -0.016367& -0.007867\\
MLE & -0.004958& -0.002026& -0.017401& -0.005958\\
LM & 0.004229& 0.002091& 0.016958& 0.005620\\
U-Statistic & 0.003600& 0.001065& 0.013052& 0.003600\\
\cline{1-5} \\ \cline{1-5}
&\multicolumn{4}{c}{n=4000}\\ \cline{1-5}
&\multicolumn{4}{c}{parameters level}\\ \cline{1-5}
Method&($\alpha=0.5$, $\beta=0.5$) &($\alpha=0.5$, $\beta=2.5$)& ($\alpha=2.5$, $\beta=0.5$)& ($\alpha=2.5$, $\beta=2.5$)\\ \cline{1-5}
GLS1 & 0.002892& 0.976e-03& 0.009939& 0.003492\\
WLS & 0.002632& 0.764e-03& 0.009614& 0.002632\\
GLS2 & -0.002260& -6.088e-04& -0.009609& -0.002260\\
MLE & -0.002372& -9.705e-04& -0.009305& -0.002372\\
LM & 0.001498& -7.164e-04& -0.007535& 0.002498\\
U-Statistic & 0.001253& 2.018e-04& -0.004859& 0.001253\\
\cline{1-5}
\end{tabular}
\end{center}
\label{tab3}
\end{table}
\begin{table}
%\tiny
\caption{Bias of scale parameter estimators for samples of large size.}
\begin{center}
\begin{tabular}{lccccccc}
\cline{1-5}
&\multicolumn{4}{c}{n=1000}\\ \cline{1-5}
&\multicolumn{4}{c}{parameters level}\\ \cline{1-5}
Method&($\alpha=0.5$, $\beta=0.5$) &($\alpha=0.5$, $\beta=2.5$)& ($\alpha=2.5$, $\beta=0.5$)& ($\alpha=2.5$, $\beta=2.5$)\\ \cline{1-5}
GLS1 & 0.006450& 0.013542& 0.006873&-0.014509\\
WLS & 0.004087& 0.011946& 0.006442& 0.013873\\
GLS2 & 0.004891& 0.011840& 0.005900& 0.011489\\
MLE & 0.005883& 0.011700& -0.005236& 0.010988\\
LM & 0.006123& 0.010598& -0.006468& 0.012319\\
U-Statistic & 0.003323& 0.007526& 0.005378& 0.009353\\
\cline{1-5} \\ \cline{1-5}
&\multicolumn{4}{c}{n=4000}\\ \cline{1-5}
&\multicolumn{4}{c}{parameters level}\\ \cline{1-5}
Method&($\alpha=0.5$, $\beta=0.5$) &($\alpha=0.5$, $\beta=2.5$)& ($\alpha=2.5$, $\beta=0.5$)& ($\alpha=2.5$, $\beta=2.5$)\\ \cline{1-5}
GLS1 & 0.001854& 0.005237& 1.968e-03&0.002354\\
WLS & 0.001263& 0.003971& 1.879e-03&0.001263\\
GLS2 & 0.001495& 0.003879& 1.807e-03&0.001495\\
MLE & 0.001382& 0.004028& 1.748e-03&0.001382\\
LM & 0.001545&-0.003890& 1.651e-03&0.001845\\
U-Statistic & 0.000552&-0.002518& 1.029e-03&0.000552\\
\cline{1-5}
\end{tabular}
\end{center}
\label{tab4}
\end{table}
\begin{table}
\small
\caption{Lifetime data (in year)}
\begin{center}
\begin{tabular}{cccccccccccc}
\cline{1-12}
30.20& 36.55& 25.11& 39.35& 27.57& 25.91& 31.50& 29.24& 18.39& 16.65& 21.85& 24.88\\
31.61& 18.74& 19.63& 28.98& 11.10& 21.66& 22.41& 26.04& 25.07& 23.48& 28.21& 25.21\\
25.12& 27.76& 23.47& 23.51& 24.39& 21.93& 37.63& 20.32& 28.17& 24.66& 30.13& 21.42\\
17.21& 19.98& 33.09& 16.04& 17.96& 19.57& 22.91& 25.69& 23.47& 16.91& 27.20& 27.23\\
\cline{1-12}
\end{tabular}
\end{center}
%\end{small}
\label{tab5}
\end{table}
\begin{table}
%\tiny
\caption{Estimation results after fitting two-parameter Weibull distribution to the lifetime data.}
\begin{center}
\begin{tabular}{lcccc}
\cline{1-5}
&\multicolumn{2}{c}{Estimated parameters}&\multicolumn{2}{c}{goodness-of-fit measures}\\ \cline{1-5}
Method&$\hat{\alpha}$& $\hat{\beta}$& KS & CVM \\ \cline{1-5}
$U$-statistic&5.1575& 26.8644& 0.0934& 0.0591\\
MLE &4.5922& 26.9452& 0.0920& 0.0713\\
WMLE &4.5141& 26.9370& 0.0906& 0.0744\\
GLS1 &4.7548& 26.9926& 0.0971& 0.0721\\
GLS2 &4.3035& 26.9788& 0.0904& 0.0926\\
WLS &4.7099& 26.6979& 0.0777& 0.0482\\
LM &4.9512& 26.9055& 0.0939& 0.0609\\
MLM &5.3119& 26.7771& 0.0889& 0.0529\\
PM & 5.9767& 25.8461& 0.0867& 0.0622\\
MM &4.9150& 26.9169& 0.0942& 0.0621\\
\cline{1-5}
\end{tabular}
\end{center}
\label{tab6}
\end{table}
\section{Conclusion}
We have introduced $U$-statistics for shape and scale parameters of two-parameter Weibull distribution. Asymptotic normality and consistency of the new estimators have been proved. Furthermore, a comprehensive Monte Carlo study have been carried out to compare the performance of the known estimators of the two-parameter Weibull distribution parameters. Since different estimators may appeal different users for different levels of sample size and parameters levels, a list of comparisons have been made in the paper for choosing desired estimator. Many facts can be concluded from this study, among them our results are the followings.
\begin{itemize}
\item for small sizes of samples the weighted least square (WLS) approach gives the best performance in the sense of bias.
\item shape estimator based on method of weighted least square (WLS) gives the best performance root of mean squared error (RMSE).
\item shape estimator based on method of percentile gives the worst performance in terms of RMSE.
\item shape and scale estimators based on $U$-statistic show the best performances in the sense of bias for large sample sizes.
\item shape and scale estimators based on generalized least square type-I (GLS1) approach show the worst performances in the sense of bias for large sample sizes.
\end{itemize}

\bibliography{teimouri}

\begin{thebibliography}{10}

\bibitem{arenas2010optimum}
J.~M. Arenas, J.~J. Narb{\'o}n, and C.~Al{\'\i}a.
\newblock Optimum adhesive thickness in structural adhesives joints using
  statistical techniques based on weibull distribution.
\newblock {\em International Journal of Adhesion and Adhesives},
  30(3):160--165, 2010.

\bibitem{buller2013statistical}
W.~Buller, B.~Wilson, L.~Van~Nieuwstadt, and J.~Ebling.
\newblock Statistical modelling of measured automotive radar reflections.
\newblock In {\em Instrumentation and Measurement Technology Conference
  (I2MTC), 2013 IEEE International}, pages 349--352. IEEE, 2013.

\bibitem{chiang2004examination}
Y.~J. Chiang, C.~Shih, C.~Lin, and Y.~Tseng.
\newblock Examination of tyre rubber cure by weibull distribution functions.
\newblock {\em International Journal of Materials and Product Technology},
  20(1-3):210--219, 2004.

\bibitem{corzo2008weibull}
O.~Corzo, N.~Bracho, A.~Pereira, and A.~V{\'a}squez.
\newblock Weibull distribution for modeling air drying of coroba slices.
\newblock {\em LWT-Food Science and Technology}, 41(10):2023--2028, 2008.

\bibitem{cousineau2009nearly}
D.~Cousineau.
\newblock Nearly unbiased estimators for the three-parameter weibull
  distribution with greater efficiency than the iterative likelihood method.
\newblock {\em British Journal of Mathematical and Statistical Psychology},
  62:167--191, 2009.

\bibitem{cran1988moment}
G.~Cran.
\newblock Moment estimators for the 3-parameter weibull distribution.
\newblock {\em IEEE Transactions on Reliability}, 37(4):360--363, 1988.

\bibitem{diaconu2009weibull}
A.~Diaconu.
\newblock Weibull distribution as a model for growth/decline in product sales.
\newblock {\em Metalurgia International}, 14:184--185, 2009.

\bibitem{dodson2006weibull}
B.~Dodson.
\newblock {\em The Weibull analysis handbook}.
\newblock ASQ Quality Press, 2006.

\bibitem{engeman1982generalized}
R.~M. Engeman and T.~J. Keefe.
\newblock On generalized least squares estimation of the weibull distribution.
\newblock {\em Communications in Statistics-Theory and Methods},
  11(19):2181--2193, 1982.

\bibitem{gebizlioglu2011comparison}
O.~L. Gebizlioglu, B.~{\c{S}}eno{\u{g}}lu, and Y.~M. Kantar.
\newblock Comparison of certain value-at-risk estimation methods for the
  two-parameter weibull loss distribution.
\newblock {\em Journal of Computational and Applied Mathematics},
  235(11):3304--3314, 2011.

\bibitem{genc2005estimation}
A.~Genc, M.~Erisoglu, A.~Pekgor, G.~Oturanc, A.~Hepbasli, and K.~Ulgen.
\newblock Estimation of wind power potential using weibull distribution.
\newblock {\em Energy Sources}, 27(9):809--822, 2005.

\bibitem{genschel2010comparison}
U.~Genschel and W.~Q. Meeker.
\newblock A comparison of maximum likelihood and median-rank regression for
  weibull estimation.
\newblock {\em Quality Engineering}, 22(4):236--255, 2010.

\bibitem{breaking}
N.~Glick.
\newblock Breaking records and breaking boards.
\newblock {\em The American Mathematical Monthly}, 85(1):2--26, 1978.

\bibitem{hassanein1971percentile}
K.~M. Hassanein.
\newblock Percentile estimators for the parameters of the weibull distribution.
\newblock {\em Biometrika}, 58(3):673--676, 1971.

\bibitem{hasumi2009weibull}
T.~Hasumi, T.~Akimoto, and Y.~Aizawa.
\newblock The weibull--log weibull distribution for interoccurrence times of
  earthquakes.
\newblock {\em Physica A: Statistical Mechanics and its Applications},
  388(4):491--498, 2009.

\bibitem{hosking1990moments}
J.~R. Hosking.
\newblock L-moments: analysis and estimation of distributions using linear
  combinations of order statistics.
\newblock {\em Journal of the royal statistical society. Series B
  (Methodological)}, pages 105--124, 1990.

\bibitem{hossain2003comparison}
A.~Hossain and W.~Zimmer.
\newblock Comparison of estimation methods for weibull parameters: complete and
  censored samples.
\newblock {\em Journal of statistical computation and simulation},
  73(2):145--153, 2003.

\bibitem{hung2001weighted}
W.-L. Hung.
\newblock Weighted least-squares estimation of the shape parameter of the
  weibull distribution.
\newblock {\em Quality and Reliability Engineering International},
  17(6):467--469, 2001.

\bibitem{jacquelin1993generalization}
J.~Jacquelin.
\newblock Generalization of the method of maximum likelihood.
\newblock {\em IEEE Transactions on Electrical Insulation}, 28(1):65--72, 1993.

\bibitem{kantar2015generalized}
Y.~M. Kantar.
\newblock Generalized least squares and weighted least squares estimation
  methods for distributional parameters.
\newblock {\em REVSTAT-Statistical Journal}, 13(3):263--282, 2015.

\bibitem{kantar2015robust}
Y.~M. Kantar and V.~Yildirim.
\newblock Robust estimation for parameters of the extended burr type iii
  distribution.
\newblock {\em Communications in Statistics-Simulation and Computation},
  44(7):1901--1930, 2015.

\bibitem{kuo2009new}
L.~Kuo-Chao, W.~Keng-Tung, C.~Chien-Song, et~al.
\newblock A new study on combustion behavior of pine sawdust characterized by
  the weibull distribution.
\newblock {\em Chinese Journal of Chemical Engineering}, 17(5):860--868, 2009.

\bibitem{lavanya2016fast}
V.~Lavanya, G.~S. Rao, and B.~Bidikar.
\newblock Fast fading mobile channel modeling for wireless communication.
\newblock {\em Procedia Computer Science}, 85:777--781, 2016.

\bibitem{lu2004note}
H.-L. Lu, C.-H. Chen, and J.-W. Wu.
\newblock A note on weighted least-squares estimation of the shape parameter of
  the weibull distribution.
\newblock {\em Quality and Reliability Engineering International},
  20(6):579--586, 2004.

\bibitem{mohan2013comparison}
C.~R. Mohan, A.~V. Rao, and G.~V. S.~R. Anjaneyulu.
\newblock Comparison of least square estimators with rank regression estimators
  of weibull distribution-a simulation study.
\newblock {\em Journal of Statistics}, 20(1), 2013.

\bibitem{muraleedharan2007modified}
G.~Muraleedharan, A.~Rao, P.~Kurup, N.~U. Nair, and M.~Sinha.
\newblock Modified weibull distribution for maximum and significant wave height
  simulation and prediction.
\newblock {\em Coastal Engineering}, 54(8):630--638, 2007.

\bibitem{nadarajah2006modified}
S.~Nadarajah and S.~Kotz.
\newblock The modified weibull distribution for asset returns.
\newblock {\em Quantitative Finance}, 6(6):449--449, 2006.

\bibitem{noga2010overview}
K.~M. Noga and B.~Pa{\l}czy{\'n}ska.
\newblock Overview of fading channel modeling.
\newblock {\em International Journal of Electronics and Telecommunications},
  56(4):339--344, 2010.

\bibitem{norman1994continuous}
L.~Norman, S.~Kotz, and N.~Balakrishnan.
\newblock Continuous univariate distributions, 1994.

\bibitem{pascual2006morchella}
P.~Pascual.
\newblock Morchella esculenta (morel) rehydration process modeling.
\newblock {\em Journal of food engineering}, 72:346--353, 2006.

\bibitem{raghunathan2002studies}
K.~Raghunathan, V.~Subramaniam, and V.~Srinivasamoorthy.
\newblock Studies on the tensile characteristics of ring and rotor yarns using
  modified weibull distribution.
\newblock 2002.

\bibitem{seki1996robust}
T.~Seki and S.~Yokoyama.
\newblock Robust parameter-estimation using the bootstrap method for the
  2-parameter weibull distribution.
\newblock {\em IEEE Transactions on Reliability}, 45(1):34--41, 1996.

\bibitem{stankova2010modeling}
T.~V. Stankova and T.~M. Zlatanov.
\newblock Modeling diameter distribution of austrian black pine (pinus nigra
  arn.) plantations: a comparison of the weibull frequency distribution
  function and percentile-based projection methods.
\newblock {\em European journal of forest research}, 129(6):1169--1179, 2010.

\bibitem{surendran2014parameter}
P.~Surendran, J.-H. Lee, S.~J. Ko, et~al.
\newblock Parameter optimization of uwb srr system performance in weibull
  clutter environment.
\newblock {\em organization}, 7(2), 2014.

\bibitem{team2014r}
R.~C. Team.
\newblock R: A language and environment for statistical computing. r foundation
  for statistical computing, vienna, austria. 2013, 2014.

\bibitem{teimouri2013comparison}
M.~Teimouri, S.~M. Hoseini, and S.~Nadarajah.
\newblock Comparison of estimation methods for the weibull distribution.
\newblock {\em Statistics}, 47(1):93--109, 2013.

\bibitem{tiryakiouglu2007estimating}
M.~Tiryakio{\u{g}}lu and D.~Hudak.
\newblock On estimating weibull modulus by the linear regression method.
\newblock {\em Journal of Materials Science}, 42(24):10173--10179, 2007.

\bibitem{van2012parameter}
J.~M. Van~Zyl and R.~Schall.
\newblock Parameter estimation through weighted least-squares rank regression
  with specific reference to the weibull and gumbel distributions.
\newblock {\em Communications in Statistics-Simulation and Computation},
  41(9):1654--1666, 2012.

\bibitem{wang1995improved}
F.~Wang and J.~B. Keats.
\newblock Improved percentile estimation for the two-parameter weibull
  distribution.
\newblock {\em Microelectronics Reliability}, 35(6):883--892, 1995.

\bibitem{wayne1982applied}
N.~Wayne.
\newblock Applied life data analysis, 1982.

\bibitem{wood2005temporal}
M.~A. Wood, B.~Gunderson, A.~Xia, X.~Zhou, V.~Padmanabhan, and K.~A.
  Ellenbogen.
\newblock Temporal patterns of ventricular tachyarrhythmia recurrences follow a
  weibull distribution.
\newblock {\em Journal of cardiovascular electrophysiology}, 16(2):181--185,
  2005.

\bibitem{zhang2008weighted}
L.~Zhang, M.~Xie, and L.~Tang.
\newblock On weighted least squares estimation for the parameters of weibull
  distribution.
\newblock In {\em Recent Advances in Reliability and Quality in Design}, pages
  57--84. Springer, 2008.

\bibitem{zhang2007study}
L.~Zhang, M.~Xie, and L.~C. Tang.
\newblock A study of two estimation approaches for parameters of weibull
  distribution based on wpp.
\newblock {\em Reliability Engineering \& System Safety}, 92(3):360--368, 2007.

\end{thebibliography}
\bibliographystyle{abbrv}
\end{document}